\documentclass[12pt]{report}
\usepackage{lipsum} 
\usepackage[margin=1in]{geometry}
\usepackage{titlesec} 
\usepackage{amsmath}
\usepackage[table,xcdraw]{xcolor}
\usepackage{pdflscape}
\usepackage[export]{adjustbox}
\usepackage{array}
\usepackage{supertabular}
\usepackage{subcaption}
\usepackage{comment}
\usepackage{color}
\usepackage{graphicx}
\usepackage{multirow}
\usepackage{lscape}
\usepackage{diagbox}
\usepackage[colorlinks,citecolor=blue,urlcolor=blue,linkcolor=blue]{hyperref}
\usepackage[title]{appendix}
\usepackage{fancyhdr}

\titleformat{\chapter}[display]
  {\normalfont\Large\bfseries\centering}{\chaptertitlename\ \thechapter}{25pt}{\LARGE}

\definecolor{lime}{HTML}{A6CE39}

\fancypagestyle{plain}{
  \fancyhf{} 
  \rhead{} 
  \lhead{Springer Studies in Computational Intelligence} 
  \lfoot{} 
}

\begin{document}

\setcounter{chapter}{0}
\chapter{Introduction to IoT}
\vspace{-20pt}
\begin{center}
    Tajkia Nuri Ananna \href{https://orcid.org/0000-0001-7385-980X}{\includegraphics[width=.125in,height=.125in,clip,keepaspectratio]{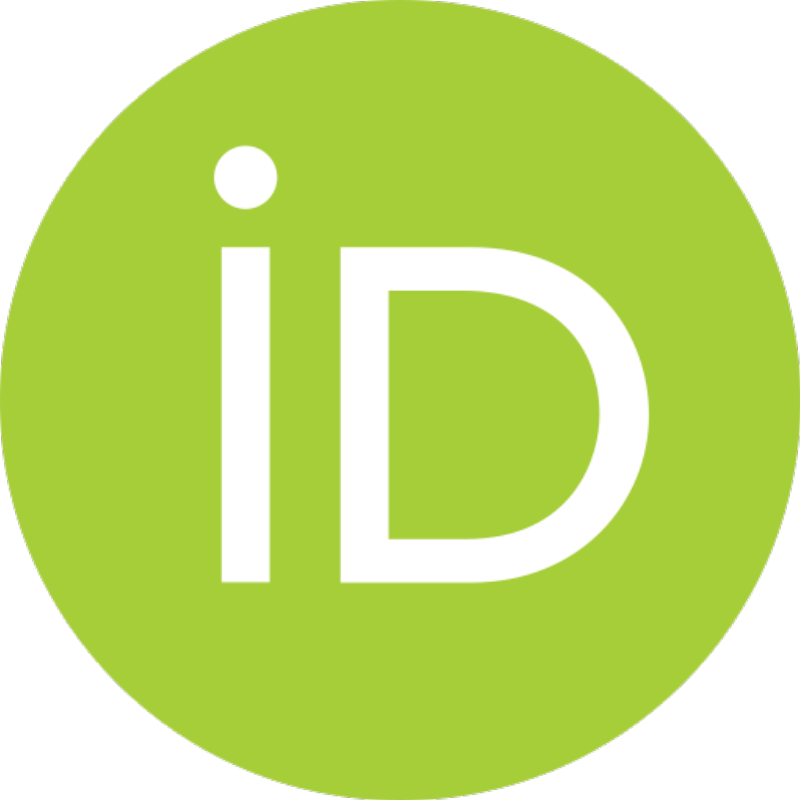}}\textsuperscript{1}, Munshi Saifuzzaman \href{https://orcid.org/0000-0002-9236-5554}{\includegraphics[width=.125in,height=.125in,clip,keepaspectratio]{images/orcid.pdf}}\textsuperscript{2} \\ 
    \vspace{10pt}
    {\fontsize{14}{16} \textsuperscript{1} Department of CSE, Metropolitan University, Sylhet 3104, Bangladesh \\
    \textsuperscript{2} Dynamic Solution innovators, Dhaka 1206, Bangladesh\\ 
    Email Address of the Corresponding Author: munshisaifuzzaman@gmail.com}
\end{center}

\section*{Abstract}

The Internet of Things has rapidly transformed the 21\textsuperscript{st} century, enhancing decision-making processes and introducing innovative consumer services such as pay-as-you-use models. The integration of smart devices and automation technologies has revolutionized every aspect of our lives, from health services to the manufacturing industry, and from the agriculture sector to mining. Alongside the positive aspects, it is also essential to recognize the significant safety, security, and trust concerns in this technological landscape. This chapter serves as a comprehensive guide for newcomers interested in the IoT domain, providing a foundation for making future contributions. Specifically, it discusses the overview, historical evolution, key characteristics, advantages, architectures, taxonomy of technologies, and existing applications in major IoT domains. In addressing prevalent issues and challenges in designing and deploying IoT applications, the chapter examines security threats across architectural layers, ethical considerations, user privacy concerns, and trust-related issues. This discussion equips researchers with a solid understanding of diverse IoT aspects, providing a comprehensive understanding of IoT technology along with insights into the extensive potential and impact of this transformative field.

\noindent\textbf{Keywords:} Architectural Layers, Ethical Considerations, Security Threats, Hardware Platforms, Trust-related Issues,  Taxonomy of technologies. 
      
\section{Introduction}
In today's ubiquitous digital landscape, the Internet has profoundly impacted global existence, marking an ongoing journey towards even more pervasive connectivity—ushering in the era of the Internet of Things (IoT). A groundbreaking invention in recent decades, IoT revolutionizes the interaction between the physical and digital realms, as defined by Vermesan et al. \cite{vermesan2022internet}. In this interconnected landscape, the digital world engages with the physical world through an array of sensors and actuators. Pena-López et al. \cite{pena2005itu} offer an expansive interpretation, characterizing IoT as a paradigm where computing and networking seamlessly integrate into virtually any object, empowering remote querying and modification. Broadly, the term "Internet of Things" describes a transformative realm where nearly every daily device is intricately linked to a network, enabling collaborative utilization for intelligent and automated tasks. The concept of IoT was first introduced by Peter T. Lewis in 1985, defining it as the fusion of individuals, processes, and technology with interconnected devices and sensors. This facilitates remote monitoring, status assessment, manipulation, and trend analysis of these devices \cite{wiki2023Internet}. The IoT journey, far from its conclusion, promises a future where diverse devices seamlessly connect to the web, reshaping human existence in unprecedented ways.

The IoT is a network of interconnected devices with sensors, actuators, processors, and various communication technologies. Sensors collect real-time data from both internal states and external surroundings, ranging from mobile phones to microwave ovens. Actuators, in turn, respond to data or commands, enabling automation and remote control of physical devices. Data collected by sensors undergoes processing either at the network edge or on central servers, with some preprocessing occurring directly in the sensors or end devices. Processed data is then transmitted to remote servers for further analysis, storage, and processing. These data form the basis for analysis, decision-making, and subsequent actions, which can be physical (e.g., adjusting a smart thermostat) or virtual (e.g., sending notifications) \cite{sethi2017internet}. The applications of IoT are extensive and diverse, impacting various aspects of our lives. From personal convenience in smart homes to healthcare and fitness innovations, IoT has the potential to influence personal, financial, physical, educational, professional, and mental aspects of individuals. In smart homes, IoT enables remote control of electrical appliances, lighting, coffee brewing, thermostat adjustments, and even hands-free operations through voice commands \cite{cook2003mavhome}. In healthcare, wearable IoT devices offer remote monitoring, allowing caregivers and healthcare professionals to provide timely assistance in emergencies. Additionally, individuals can use wearable devices to track sleep patterns, physical activity, and overall fitness \cite{dalal2020internet}. These examples only scratch the surface of IoT's broad application landscape, indicating the exciting possibilities and challenges that researchers are exploring for the future.


The IoT has the potential to transform how people interact with technology, providing greater convenience, efficiency, and personalization in daily life. Despite its transformative impact, IoT faces challenges. The sheer number of devices and the substantial data generated pose significant challenges, with a projected 41.6 billion IoT devices producing 79.4 zettabytes of data by 2025 \cite{dellinternet}. Addressing this requires scalable architectures and enhanced processing capabilities. Moreover, IoT heavily relies on wireless communication, leading to challenges such as distortion and unreliability in geographically dispersed locations. Ensuring dependable data transmission becomes a pivotal challenge, emphasizing the critical role of communication technologies in the IoT landscape. Beyond technical hurdles, various general and domain-specific challenges are crucial for the success of IoT. Identifying and addressing these multifaceted challenges collectively is essential to unlocking the full potential of IoT and overcoming obstacles for broader adoption and integration into our lives.

\textbf{Motivation of this chapter:} IoT is not a novel concept; researchers have been exploring this field for decades. Consequently, the question arises: what sets this chapter apart from others, and why should readers invest their time in exploring the basics presented here? While there are several book chapters introducing IoT, many share outdated concepts, lacking the latest insights from the ongoing exploration of this dynamic field. For instance, in \cite{kramp2013introduction}, only IoT applications are discussed, and in \cite{lakhwani2020internet}, authors overlook major domains, including IoT applications, advantages, challenges, and technologies. Notably, Nagaraj et al. \cite{nagaraj2021introduction} present a well-structured discussion covering technologies, architectures, applications, and challenges. However, their discussion is limited to these aspects, with a less extensive exploration of the application section. Therefore, the need arises for a comprehensive chapter that covers all aspects and incorporates recent additions in the IoT field. This chapter goes beyond the basics, encompassing fundamental components, characteristics, and advantages. It delves into architectures, provides a taxonomy of technologies used in IoT, explores a significant number of applications across diverse domains, and addresses ethical considerations as well as legal and regulatory issues. In essence, our chapter serves as a holistic guide, covering the most significant facets of IoT, from foundational principles to emerging research challenges and future directions. Readers are encouraged to explore this chapter for a thorough and up-to-date understanding of the evolving landscape of IoT.


\textbf{Contributions:} The contributions of this chapter can be summarized as follows:
\begin{enumerate}
    \item This chapter provides an introductory overview of the IoT with the aim of assisting future contributors. It assesses the benefits, generic architecture, key technologies underpinning IoT, and its diverse applications across various domains.
    
    \item Recent studies have been examined to facilitate this assessment. Additionally, this chapter encompasses a taxonomy of IoT technologies, including extensive coverage of field communication.
    
    \item In conclusion, this chapter offers a comprehensive discussion of the open research challenges, ethical considerations, as well as the legal and regulatory aspects of IoT.
\end{enumerate}

\begin{table}[b]
\centering
\caption{List of abbreviations}
\label{tab:abbreviations}
\resizebox{\columnwidth}{!}{%
\begin{tabular}{l|l||l|l||l|l}
\hline
\rowcolor[gray]{.9} \textbf{Name} & \textbf{Abbreviation} & \textbf{Name} & \textbf{Abbreviation} & \textbf{Name} & \textbf{Abbreviation} \\ \hline \hline
FDMA & Frequency Division Multiple Access & LBT & Listen Before Talk & CDMA & Code Division Multiple Access \\ 
\rowcolor[gray]{.96} FHSS & Frequency Hopping Spread Spectrum & CSS & Chirp Spread Spectrum & OOK & On-Off Keying \\ 
TDMA & Time Division Multiple Access & GSM & \begin{tabular}[c]{@{}l@{}}Global System for Mobile \\ Communications\end{tabular} & PPM & Pulse Position Modulation \\ 
\rowcolor[gray]{.96}GFSK & Gaussian Frequency Shift Keying & UMTS & \begin{tabular}[c]{@{}l@{}}Universal Mobile Telecommunications\\  System\end{tabular} & OFDM & \begin{tabular}[c]{@{}l@{}}Orthogonal Frequency Division\\  Multiplexing\end{tabular} \\ 
DWPSK & \begin{tabular}[c]{@{}l@{}}Differential Quadrature Phase Shift\\  Keying\end{tabular} & GMSK & Gaussian Minimum Shift Keying & TDD & Time Division Duplex \\ 
\rowcolor[gray]{.96}CSMA/CA & \begin{tabular}[c]{@{}l@{}}Carrier Sense Multiple Access with\\  Collision Avoidance\end{tabular} & LTE-A & Long-Term Evolution Advanced & FDD & Frequency Division Duplex \\ 
BPSK & Binary Phase Shift Keying & 3GPP & 3rd Generation Partnership Project & ALHOA & Adaptive Link Hopping On Air \\ 
\rowcolor[gray]{.96}CSMA/CD & \begin{tabular}[c]{@{}l@{}}Carrier Sense Multiple Access with\\  Collision Detection\end{tabular} & WCDMA & \begin{tabular}[c]{@{}l@{}}Wideband Code Division Multiple\\  Access\end{tabular} & DBPSK & \begin{tabular}[c]{@{}l@{}}Differential Binary Phase Shift\\  Keying\end{tabular} \\ 
O-QPSK & Offset Quadrature Phase Shift Keying & OFDMA & \begin{tabular}[c]{@{}l@{}}Orthogonal Frequency Division \\ Multiple Access\end{tabular} & QAM & Quadrature Amplitude Modulation \\ 
\rowcolor[gray]{.96}QPSK & Quadrature Phase Shift Keying & SC-FDMA & \begin{tabular}[c]{@{}l@{}}Single Carrier Frequency Division\\  Multiple Access\end{tabular} & CP-OFDM & \begin{tabular}[c]{@{}l@{}}Cyclic Prefix Orthogonal Frequency\\  Division Multiplexing\end{tabular} \\ \hline
\end{tabular}%
}
\end{table}

\textbf{Chapter organization:} It begins with an exploration of the fundamentals in Section \ref{sec:fundamentals}, covering the history \ref{sec:history}, components \ref{subsec:componentsOfIoT}, and characteristics \ref{sec:char}. In Section \ref{sec: adv_of_IoT}, the advantages of IoT are examined. By introducing a generic architecture in Section \ref{sec:architecture}, the taxonomy of technologies that underpin IoT operations are discussed in detail in Section \ref{sec:technologies}. Furthermore, Section \ref{sec:applications} explores the extensive array of application domains of IoT, with a focus on multiple promising research works within each domain, aiming to provide a comprehensive understanding of its real-world applications. It concludes in Section \ref{sec:challenges} by addressing the key challenges that IoT faces today, considering insights from the reviewed literature, and sheds light on potential future scopes and developments in the ever-evolving realm of IoT technology.
Table \ref{tab:abbreviations} represents some abbreviations mentioned in this chapter.

\section{Fundamentals of IoT}
\label{sec:fundamentals}
At its core, the IoT revolves around the concept of connecting everyday objects and devices to the internet, enabling them to communicate, collect data, and perform actions autonomously or in response to commands. This basic concept involves three key elements: sensors and actuators that gather and interact with data, a network infrastructure for data transmission, and cloud-based platforms for data storage, processing, and analysis. By interconnecting these elements, IoT empowers us to enhance efficiency, gain real-time insights, and create smart, responsive systems that impact various aspects of our lives, from smart homes and cities to industries and healthcare. The evolution of IoT, which has unfolded over several decades, began with the concept of smart objects. Since then, IoT has undergone numerous groundbreaking transformations that have left the world astounded, thanks to its unique and convenient characteristics. The advantages brought by this technology are unquantifiable, as it touches nearly every aspect of people's lives, simplifying and improving them. This section comprehensively discusses these fundamental IoT concepts, covering the following topics: \textit{history}, \textit{components}, \textit{characteristics}.

\subsection{Historical Development}
\label{sec:history}
Since the invention of the first landline, the telegraph, in the 1830s and 1840s, machines have been instrumental in facilitating direct communication. A significant step towards the IoT occurred on June 3, 1900, with the first radio voice transmission, often referred to as ``wireless telegraphy''. This paved the way for IoT. The development of computers, which began in the 1950s, is another crucial aspect of IoT.

In 1962, the Internet, a fundamental component of IoT, started as a DARPA\footnote{The Defense Advanced Research Projects Agency is a research and development agency of the United States Department of Defense responsible for the development of emerging technologies for use by the military.} project. A group of renowned researchers initiated efforts to connect computers and systems. By 1969, DARPA had evolved into ARPANET\footnote{The Advanced Research Projects Agency Network (ARPANET) was the first wide-area packet-switched network with distributed control and one of the first computer networks to implement the TCP/IP protocol suite. Both technologies became the technical foundation of the Internet.}, a precursor to today's Internet \cite{greengard2023Internet}.

While the term ``Internet of Things'' is relatively new, the concept of integrating computers and networks to monitor and manage devices has a rich history spanning decades. In the late 1970s, various stakeholders, including businesses, governments, and consumers, began exploring ways to connect personal computers (PCs) and other machinery. This led to the practical use of systems for remotely monitoring electrical grid meters via telephone lines during that era \cite{wiki2023Machine}.

In the 1980s, there was a growing interest in enhancing physical objects with sensors and intelligence. Commercial service providers began supporting public access to ARPANET, an early precursor to the modern Internet, during this time. Satellites and landlines played pivotal roles in establishing the foundational communication infrastructure for the emerging IoT. The concept of a network of smart devices was initially explored as early as 1982 when a Coca-Cola vending machine\footnote{The "Only" Coke Machine on the Internet. Available at: \href{https://www.cs.cmu.edu/~coke/history_long.txt}{Carnegie Mellon University}} at Carnegie Mellon University was modified to connect to ARPANET. This allowed local programmers to remotely monitor the vending machine's contents, ensuring drinks were available and cold before making a purchase. However, the technology was challenging to manage, and progress in this field was limited during that period \cite{humanity2023IoT}. In parallel, during the 1980s, local area networks (LANs) gained popularity and proved effective for real-time communication and document sharing among groups of PCs.

In the 1990s, advancements in wireless technology set the stage for the widespread adoption of "machine-to-machine" (M2M) solutions in enterprise and industrial contexts, particularly for equipment monitoring and operation. However, many of these early M2M solutions relied on closed, purpose-built networks and proprietary or industry-specific standards rather than utilizing Internet Protocol (IP)-based networks and Internet standards \cite{rose2015internet}. By the mid-1990s, the Internet had expanded its global reach, offering new possibilities for researchers and technologists to explore ways to enhance connections between humans and machines. A significant milestone in this journey was the creation of the first Internet-connected 'device' by John Romkey—an IP-enabled toaster that could be controlled over the Internet. This innovative toaster was showcased at an Internet conference in 1990, marking an early example of the IoT in action \cite{romkey2016toast}.

In 1991, Mark Weiser's paper ``The Computer of the 21st Century'' \cite{weiser1991computer} and academic events like UbiComp and PerCom shaped the contemporary IoT vision \cite{rose2015internet}. Global Positioning System (GPS) became a reality in early 1993 with the Department of Defense establishing a stable system of 24 satellites. Privately owned commercial satellites soon followed, enhancing IIoT functionality \cite{wiki2023Internet}. In early 1994, Reza Raji introduced the IoT concept in IEEE Spectrum, describing it as ``moving small data packets to integrate and automate from home appliances to entire factories'' \cite{raji1994Smart}. Later that year, Steve Mann invented the near-real-time WearCam, powered by a 64-processor setup. Between 1993 and 1997, companies proposed IoT solutions, including Microsoft's ``at Work'' and Novell's NEST. Momentum grew as Bill Joy introduced device-to-device communication in his "Six Web" framework at the 1999 World Economic Forum \cite{wiki2023Internet}.

The term ``Internet of Things'' was coined by Peter T. Lewis in a 1985 speech during the Congressional Black Caucus Foundation's 15th Annual Legislative Weekend in Washington, D.C. Lewis defined IoT as the integration of people, processes, and technology with connectable devices and sensors for remote monitoring, status assessment, manipulation, and trend evaluation related to these devices \cite{wiki2023Internet}. In 1997, Paul Saffo described sensors and their future roles. British technologist Kevin Ashton, while serving as the executive director of the Auto-ID Center at MIT, independently coined the term ``Internet of Things''. During his time at Procter and Gamble, he explored radio-frequency identification (RFID), a technology framework enabling physical devices to connect via microchips and wireless signals. In the same year, they developed a global RFID-based item identification system \cite{rose2015internet}.

In 1999, Kevin Ashton was the first to describe the IoT and proposed the name ``Internet of Things'' during a presentation for Procter and Gamble. He believed RFID technology, primarily designed for inventory tracking, was a significant prerequisite for the IoT, allowing computers to efficiently manage and monitor individual objects. The concept of tagging objects has been realized through technologies like digital watermarking, barcodes, and QR codes, used for identification and tracking purposes \cite{foote2022Breif}. Subsequent technological advancements, including the proliferation of smartphones, cloud computing, improved processing power, and enhanced software algorithms, along with the availability of sophisticated sensors capable of measuring various parameters, laid the foundation for robust data collection, storage, and processing for the IoT's growth.

As a significant step forward in commercializing IoT, LG announced plans to launch a smart refrigerator capable of autonomously managing its contents in 2000. Walmart and the US Department of Defense pioneered inventory tracking using RFID and the IoT in $2002-2003$. RFID gained prominence in the US Army's Savi program in 2003, and Walmart expanded its RFID usage worldwide that same year. In 2004, Cornelius ``Pete'' Peterson, CEO of NetSilicon, predicted that IoT devices would dominate the next era of information technology, particularly in fields like medical devices and industrial controls \cite{wiki2023Internet}. In 2005, numerous articles in mainstream newspapers such as The Guardian, Scientific American, and The Boston Globe discussed IoT's future direction.

The IPSO Alliance was founded in 2008 to promote the use of IP in networks of ``smart objects'', while the FCC allowed the use of the ``white'' label in 2008. Google initiated the development of autonomous cars in 2009, and in 2011, Google's Nest smart thermostat entered the market, enabling remote heating management. In June 2012, major Internet service providers and web-based companies agreed to expand the global Internet's address space by enabling IPv6 for their services and products, a significant step towards a viable IoT. This led to substantial growth and interest in the field. IT giants like Cisco, IBM, and Ericsson later took numerous educational and commercial initiatives related to IoT. Cisco Systems estimated that the IoT was "born" between 2008 and 2009, with the things/people ratio growing from 0.08 in 2003 to 1.84 in 2010 \cite{foote2022Breif}.

\subsection{Components}
\label{subsec:componentsOfIoT}
The IoT consists of several key components that serve as the essential building blocks for constructing an IoT system. This section provides an in-depth exploration of the principal components of the IoT. IoT comprises three main components: (1) sensors/devices, and actuators; (2) storage and data analytics; and (3) interpretation and visualization tools. Each of these is further categorized into various subcomponents.

\vspace{0.5cm}
\noindent\textit{Sensors/Devices and Actuators}\\
\begin{enumerate}
    \item Sensors/Devices: Sensors play a crucial and essential role within an IoT system. Given that IoT operates by gathering data from the surrounding environment, it is necessary for all IoT applications to incorporate one or more sensors to meet this need. A defining characteristic of IoT devices is their context awareness, which is made possible through the utilization of sensor technology. Sensors are not only compact and cost-effective but also energy-efficient. However, they are subject to limitations such as battery capacity and ease of deployment. \cite{sethi2017internet}. An overview of various types of sensors has been provided below.
    \begin{enumerate}
        \item Mobile-based Sensors\\ Smartphones, which are widespread and commonly used, are equipped with various sensors. Given their extensive usage, researchers are exploring the potential of using smartphones as integral components in building smart IoT solutions. These applications can harness sensor data from smartphones to generate valuable insights and outcomes. Some of the general sensors found in smartphones include an accelerometer, gyroscope, GPS, magnetometer, light sensor, and proximity sensor \cite{schmidt2001build}. Certain smartphones, like the Samsung Galaxy S4, come equipped with extra sensors, including a thermometer, barometer, and humidity sensor \cite{sethi2017internet}.

        \item Medical Sensors\\The healthcare industry is one of the most influential fields where innovation and IoT have paved the way. Wearable devices and sensors have facilitated remote monitoring for physicians and enabled researchers to collect data continuously and in real-time.
        These devices come in different forms, such as wristbands, smartwatches, and monitoring patches. Smartwatches and fitness trackers, known for their versatility, have gained popularity among consumers. Likewise, monitoring patches have emerged as a valuable asset to the healthcare sector by enabling remote treatment for patients.
        \item Neural Sensors\\Neural sensors play a crucial role in comprehending the workings of human neurons by enabling us to decode brain signals, assess the brain's current state, and, when necessary, optimize it for improved focus and attention. This practice is commonly referred to as \textit{Neurofeedback} \cite{gruzelier2014eeg}.
        \item Environmental and Chemical Sensors\\
       While conventional tools manage parameters such as temperature and pressure, specialized environmental sensors play a crucial role in evaluating air quality. These sensors detect gases and particulate matter \cite{sekhar2010chemical}, while also measuring factors like temperature, humidity, pressure, and pollution. Besides, chemical sensors play a crucial role in detecting both chemical and biochemical substances. Among the innovative technologies available are the electronic nose (e-nose) and electronic tongue (e-tongue), which rely on pattern recognition to sense chemicals based on odor and taste. These sensors find valuable applications in smart cities for monitoring pollution levels \cite{manna2014vehicular}.
\item Radio-Frequency Identification (RFID)\\
RFID technology, which serves as sensors, finds widespread use in various IoT applications. For instance, it is employed for tracking products within extensive inventories or monitoring items within large retail stores.


    \end{enumerate}
    \item Actuators: Actuators hold a crucial role and operate in direct contrast to sensors. They transform energy into physical motion and are typically positioned on the outer periphery of a system. Take, for instance, a scenario involving a smart home system that incorporates numerous sensors and actuators. In this setup, the actuators receive signals from the sensors and, depending on the context, carry out actions such as locking or unlocking doors, toggling lights or electrical devices on or off, regulating the house's temperature, or setting alarms for emergencies. Essentially, actuators respond to and execute commands based on the signals they receive from sensors or other devices.

\end{enumerate}

\vspace{0.5cm}
\noindent\textit{Storage and Data Analytics}\\
Another crucial aspect of IoT is the management of the substantial volume of data generated and exchanged by IoT devices continuously. Storing this data presents a significant challenge within IoT networks. Furthermore, the data collected from these devices must undergo filtering, processing, and analysis to enable the effective functioning of the IoT system. In this process, gateways, cloud services, and analytics collaborate to handle data storage and processing tasks.
\begin{enumerate}
    \item Gateway: Gateways, designed to simplify the IoT system, function as the intermediate medium of communication between the devices and the central cloud system. The major functionalities of IoT gateways are listed below:
    \begin{enumerate}
    \item \textit{Data Preprocessing}\\The IoT gateway serves as an intermediary between sensor devices and the central cloud, conducting basic data analytics before forwarding information directly to the cloud. This layer performs the tasks of local data filtering, cleaning, preprocessing, and protocol translation. During this process, it may also aggregate, remove duplicates, or summarize the data to enhance response times and lower transmission costs \cite{AlexanderIoT}.
    \item \textit{Data Acquisition}\\ In this layer, data is collected from multiple sources, converted into the desired format, and then transferred to the processing layers. The role of the gateway in this stage is to provide secure connectivity between IoT devices and processing structures \cite{checkWhat}.
        \item \textit{Data forwarding and Temporary Storage}\\
        The primary role of the gateway is to ensure secure data transfer between the sensor layer and the central cloud \cite{daviteq2020IoT}. Additionally, this layer serves as the temporary storage repository for the collected data. 
        \item \textit{Device Management}\\ This layer facilitates real-time device configuration, allowing adjustments to device statuses, operational modes, error acknowledgments, and more \cite{daviteq2020IoT}.
        \item \textit{Diagnostics} \\ The IoT gateway identifies errors and faults within the entire technology layer, including self-diagnostics for the IoT gateway itself \cite{daviteq2020IoT}.
    \end{enumerate}
    \item Cloud: The cloud serves as the central hub of an IoT network, taking on pivotal roles in data processing, storage, and management. 
   The key characteristics of the cloud include the ability to store and process extensive data generated by devices, scalability to effortlessly handle thousands of devices, flexibility by allowing devices to be added or removed as needed without requiring a complete system reconfiguration, supervision and management by the cloud service provider, and cost-effectiveness.
    
    While cloud services are not mandatory for IoT, the recent shift toward edge and fog computing empowers local data processing. Nevertheless, the cloud is incorporated into the system for its scalability. storage and cost-effective service provision \cite{gubbi2013internet}. Furthermore, cloud-based services offer security functionalities such as encryption and authentication while enabling remote access and control of IoT devices.
    \item  Analytics: This represents one of the most intricate and vital layers within IoT. It involves the analysis of data, generating valuable insights through the application of diverse machine learning (ML) algorithms and statistical analysis techniques. Numerous applications of analytics in IoT encompass anomaly detection, environmental monitoring, energy management, smart cities, and agriculture \cite{tisha2023What}.
\end{enumerate}

\begin{figure}[b]
    \centering
    \includegraphics[width=0.7\textwidth]{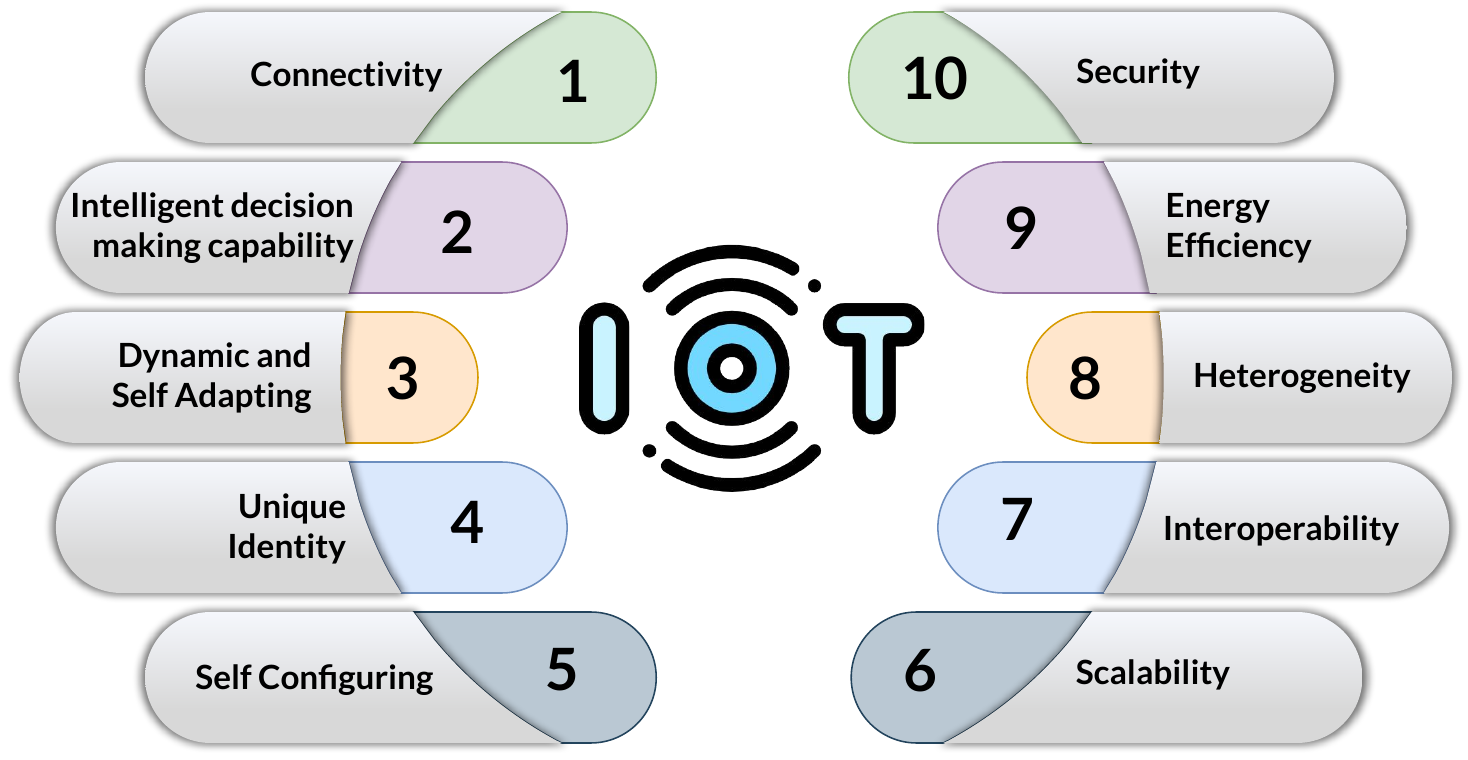}
    \caption{IoT characteristics}
    \label{iot_characteristics}
\end{figure}

\vspace{0.5cm}
\noindent\textit{Interpretation and Visualization Tools}\\
This segment essentially serves as the user interface (UI). The UI offers a platform for users to interact directly with the application or system, facilitating communication. A user interface is not always reliant on a screen. For instance, a TV remote utilizes a user interface comprising multiple buttons, while devices like the Amazon Echo respond to voice commands for control. \textit{Receiving Automatic Notification}, \textit{Monitoring Information Proactively}, and \textit{Controlling the System Remotely} are some common examples for user interfaces in IoT systems \cite{anniMachine}. 

\subsection{Characteristics}
\label{sec:char}
This segment essentially serves as the user interface (UI). The UI offers a platform for users to interact directly with the application or system, facilitating communication. A user interface is not always reliant on a screen. For instance, a TV remote utilizes a user interface comprising multiple buttons, while devices like the Amazon Echo respond to voice commands for control. Receiving automatic notifications, monitoring information proactively, and controlling the system remotely are some common examples of user interfaces in IoT systems \cite{anniMachine}. The IoT characteristics discussed in this chapter are illustrated in Figure \ref{iot_characteristics}.


\begin{enumerate}
    \item \textit{Connectivity}: Connectivity is the most vital requirement of IoT. The main aspect of IoT is a network with millions of devices connected to each other. The connectivity remains constant, allowing anyone from anywhere to connect to the IoT network at any given moment.
    
    \item  \textit{Intelligent decision making capability}: The extraction of knowledge from the data generated is highly significant. Consider a sensor that produces data; however, the true value of that data lies in its proper interpretation. This represents a crucial aspect of IoT, wherein IoT devices possess the capability to transform raw data collected by sensors into meaningful information and make decisions based on it.
    
    \item \textit{Dynamic and Self Adapting}: IoT devices should have the ability to adapt to changes in context, their surrounding environment, and the existing situation. For instance, within a surveillance system, cameras are capable of switching between day and night modes or adjusting their resolution in response to motion detection, demonstrating their adaptability.
    
    \item \textit{Unique Identity}: Every IoT device should have a unique identity and unique identifier. IoT device interfaces enable users to inquire about device information, monitor their status, and remotely manage them. Having a distinct identity is essential to empower users to safeguard their devices, whether through password protection or alternative security measures \cite{ruddhi2023What}. 
    
    \item \textit{Self Configuring}: IoT devices possess the capability to autonomously update their systems in response to the situation, eliminating the need for user intervention. Moreover, they exhibit flexibility in network management, allowing new devices to seamlessly join the network and permitting any device to depart from the network at any time.
    
    \item \textit{Scalability}: The IoT network is experiencing a continual growth in the number of connected devices, resulting in a substantial and continuous generation of data. Consequently, scalability emerges as the foremost feature of any IoT system.

    \begin{figure}[t]
        \centering
        \includegraphics[width=0.75\textwidth]{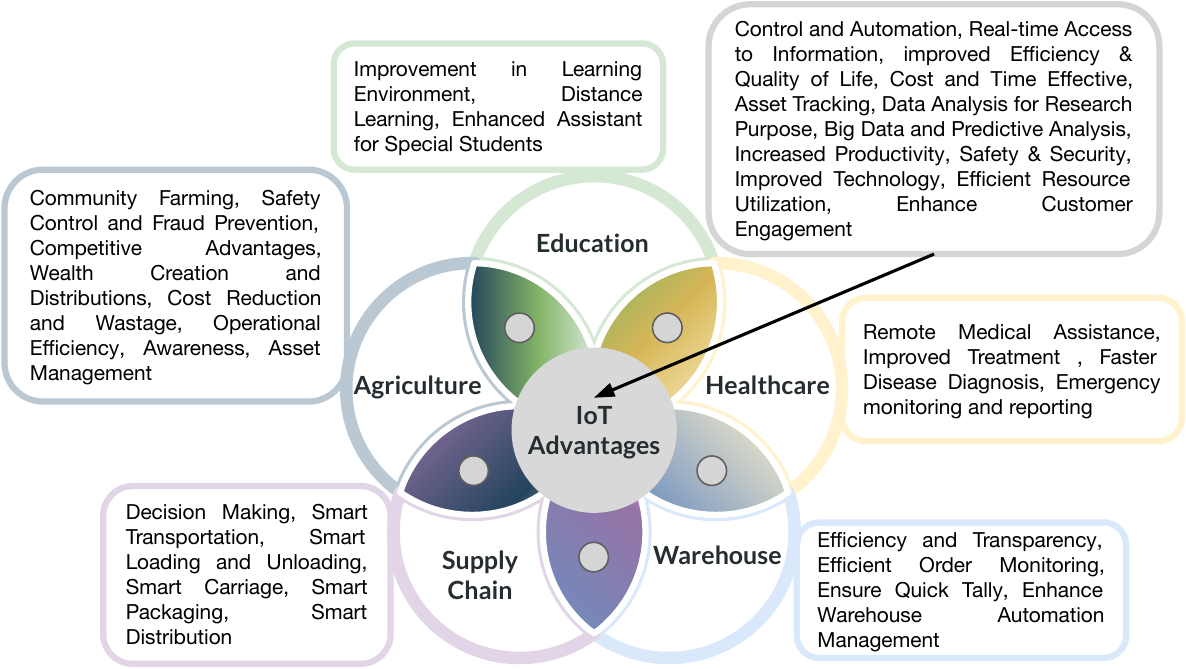}
        \caption{Advantages of IoT}
        \label{fig:advantages}
    \end{figure}
    
    \item \textit{Interoperability}: IoT devices rely on standardized protocols and technologies to guarantee seamless communication among themselves and with other systems. Interoperability represents a fundamental block of the IoT, signifying the capacity for various IoT devices and systems to interact and share data, irrespective of the underlying technology or the manufacturer. Hence, IoT devices employ standardized protocols, data formats, and technologies to uphold interoperability. 
    
    \item \textit{Heterogeneity}: The devices within an IoT network exhibit heterogeneity, showcasing the network's capability to accommodate diverse elements.
    
    \item \textit{Energy Efficiency}: Energy efficiency stands out as a significant characteristic of IoT. Numerous IoT devices are purposefully crafted to minimize energy consumption and create low-power devices. Furthermore, various IoT approaches are tailored to optimize power usage, such as opting for fog/edge computing over cloud computing to decrease bandwidth requirements and reduce power consumption when constructing an IoT system \cite{kuyoro2015internet}.
    
    \item \textit{Security}: The presence of millions of internet-connected devices and the vast amount of data generated underscore the vulnerability of the IoT network to security threats. Hence, safety and security emerge as pivotal characteristics of the IoT. Ensuring safety is of paramount importance to preserve the effectiveness of IoT's advantages, including efficiency and novel experiences. 
\end{enumerate}
\section{Advantages of IoT}
\label{sec: adv_of_IoT}
The amount of benefits IoT offers to people's lives is the main reason IoT is becoming more popular by the day. IoT solutions are developed and invented to make people's lives easier and more convenient. IoT technology affects almost every field, including healthcare, education, and business. This section elaborates on the most visible benefits of IoT in people's daily lives \cite{zubovich2023Advantages, laghari2021review}, and Fig. \ref{fig:advantages} provides a visual overview.

\begin{enumerate}
    \item \textit{Efficient Data Collection}:
    IoT-based data collection has been particularly beneficial in sectors such as healthcare and finance \cite{gimpel2020bringing}. A straightforward illustration of efficient data collection is the integration of IoT in the retail sector. Internet-connected tags can provide data on purchase decisions and sales trends, whether weekly or monthly. This enhanced data gathering can improve inventory management and reveal valuable insights into customer behavior, ultimately contributing to the prosperity of the business. 

\item \textit{Control and Automation}: IoT has provided its customers with a more convenient lifestyle and allows them to control their daily activities with the touch of a button. One simple yet remarkable example is the smart bulb, which can be controlled without even touching the switches; the user can simply turn off or on the light remotely. Not only can the light bulb, coffee maker, or any other electrical device in the home be controlled with the tip of the device, but they can also be controlled using a voice command when they are linked to Google's or Amazon's voice assistants.

\item \textit{Real-time Access to Information}: IoT devices offer immediate access to information, proving invaluable across healthcare, business, and everyday applications. A prime illustration of the advantages of real-time data access is within the healthcare sector. Physicians can access patient data in real-time, enabling continuous health monitoring. This capability becomes particularly crucial in delivering emergency medical assistance swiftly when unexpected health issues arise.

\item \textit{Improved Efficiency}: IoT systems operate autonomously, which is a valuable asset in various domains. Reduced human intervention leads to increased efficiency and decreased labor reliance. For instance, a company with a fleet of delivery vehicles can effortlessly track their real-time locations, eliminating the need for manual employee involvement in this task.

\item \textit{Improved Quality of Life}: The advent of IoT has significantly improved the lives of its users in numerous ways. Real-time health monitoring, including devices like blood pressure monitors and fitness trackers, empowers users to maintain their well-being effectively. Smart homes offer a stress-free and effortless lifestyle. These advantages extend beyond individuals and can benefit entire industries or communities. Smart devices, linked not only to intelligent traffic lights but also to road safety monitors and toll gates, can provide drivers with real-time information about road conditions on their route.

\item \textit{Cost and Time Effective}: IoT minimizes human effort and relies heavily on real-time data transmission, leading to time savings. For example, real-time patient monitoring benefits both patients and doctors by eliminating the need for physical meetings, thus saving time for both parties. IoT aids businesses in streamlining their workflows by offering valuable insights and real-time information, resulting in cost reductions. In addition to businesses, individuals can reduce their everyday expenses through the use of IoT. 


\item \textit{Asset Tracking}:
This process involves tracking products within a business or logistics management system. Manual asset tracking is labor-intensive and time-consuming, but it can be streamlined through the application of IoT technologies like barcodes and RFID tags. These technologies allow for remote monitoring of goods and provide stakeholders with information about any faults or problems in real time \cite{khan2022application}.

\item \textit{Data Analysis for Research Purpose}: The enormous amount of data collected from IoT devices has become a blessing for researchers in various fields like healthcare, education, business, etc. The healthcare researchers can use the data collected via biosensors to invent cures and vaccines for disease; the finance industry can use the data to understand trends and improve customer experience; the super shops can analyze customer behavior and improve their businesses; and so on.

\item \textit{Big Data and Predictive Analysis}: ``Big data'' has been a widely recognized term in the world long before the emergence of IoT. It involves the collection and analysis of massive volumes of data. One of the primary objectives of IoT is to amass data from diverse sources, sending this information back to systems for analysis. Effective analysis of big data can yield valuable insights, spanning from stock market predictions to understanding customer behavior, thereby enhancing the business landscape.

\item \textit{Increased Productivity}: IoT utilization in both industry and homes has the potential to boost productivity significantly. For instance, in a smart home, users can streamline various household tasks using voice commands, enabling efficient multitasking. Similarly, in a business setting, analyzing customer behavior can enhance customer satisfaction, ultimately contributing to the prosperity of the enterprise. As an example, 46\% of businesses that embraced IoT strategies saw improvements in efficiency, even though only 29\% initially anticipated such improvements \cite{impactmybiz2022Benefits}. In the healthcare sector, doctors can offer more extensive services to their patients if they do not need to make physical visits to each patient individually.

\item \textit{Safety and Security}: The incorporation of IoT offers users a means of security not only in their homes but also in their businesses, schools, offices, and virtually anywhere. Individuals can remotely monitor their valuable assets, such as vehicles and other important items. Parents can even keep track of their children's whereabouts from their workplaces, providing peace of mind. IoT allows for vehicle tracking and the setup of alert systems in case of unusual incidents. Financial companies and banks can enhance the security of their confidential rooms or vehicles by utilizing IoT \cite{kuyoro2015internet}.


\item  \textit{Enhance Customer Engagement}: The IoT offers several avenues for enhancing customer engagement. It achieves this by leveraging valuable customer data, personalizing experiences, improving convenience, and enabling real-time interactions.

 \item \textit{Efficient Resource Utilization}:IoT facilitates efficient resource utilization through various mechanisms and capabilities. Its ability to collect and monitor real-time information allows organizations to track the status of their resources, such as equipment and machinery, enabling the identification of inefficiencies. Predictive maintenance powered by IoT in the industrial sector can anticipate machinery failures, reducing downtime and optimizing resource allocation for maintenance.

 \item \textit{Improved Technology}: The innovation brought about by IoT leads to the creation of newer and more advanced technologies on the market. For instance, consider the scenario of an air conditioner that was initially controlled manually with a remote. With the advent of IoT, users can now operate it using voice commands or control it remotely. When such innovations hit the market, they spark competition, driving the development of improved solutions based on user feedback. This cycle of innovation in response to IoT advancements contributes to the continuous progression of technology \cite{gimpel2020bringing}.
\end{enumerate}

\section{IoT Architecture}
\label{sec:architecture}
Architecture refers to a structured framework depicting the tangible components of a network, their operational arrangement and setup, underlying principles and structures, as well as the manner in which data is organized and utilized within its functioning. IoT architecture comprises a collection of devices, sensors, actuators, end users, cloud services, and most importantly, various communication layers and IoT protocols. All IoT systems inherently follow the generic three-layer architecture. However, based on necessities or specific application requirements, the generic model can be modified by adding extra layers, thus forming four- or five-layer architectures \cite{sethi2017internet}.

\begin{figure}[t]
    \centering
    \includegraphics[width=0.65\textwidth]{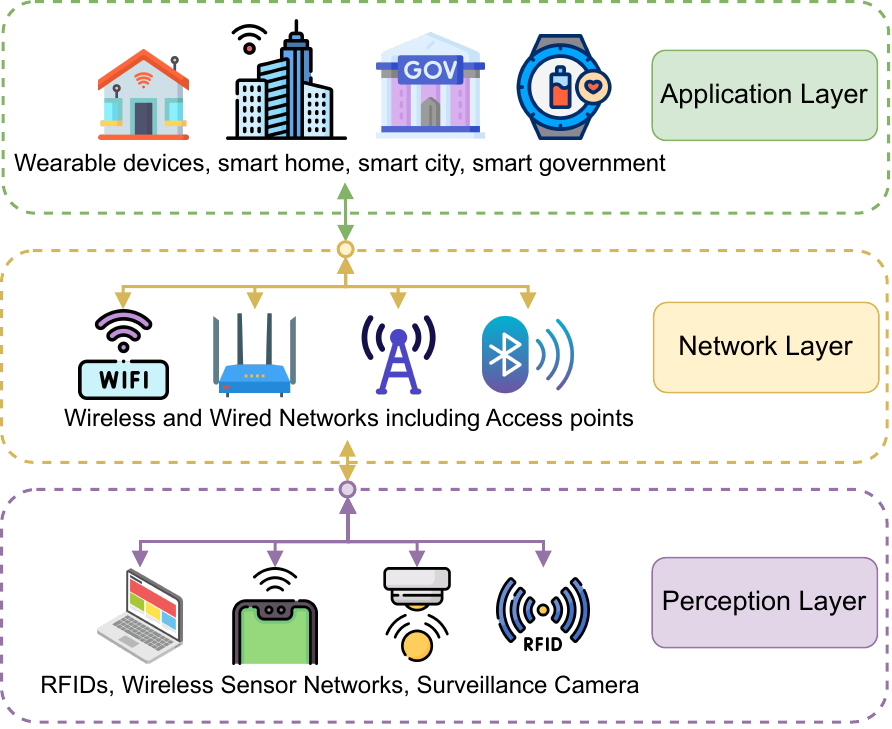}
    \caption{Generic architecture of IoT}
    \label{fig:arch_three_layer}
\end{figure}

The IoT architecture can be divided in two ways: (1) layered architecture and (2) domain-specific architecture. In this section, we have presented a generic architecture of IoT. IoT follows a layered architecture, which refers to the structured framework used to design and organize the various components and functions of an IoT system. This architecture is composed of multiple layers, each with its own specific role and responsibilities, allowing for efficient communication, data processing, and management within the IoT ecosystem. The typical layers in the IoT architecture include the perception layer, network layer, and application layer. These layers collaborate to ensure the smooth operation of IoT devices, data transmission, and the provision of IoT services to end users.

According to this generalized architecture, also known as the three-layer architecture, the IoT system is divided into three layers \cite{atzori2012social}, namely (i) the application layer, (ii) the network layer, and (iii) the perception layer. Every one of these layers possesses inherent security challenges \cite{mahmoud2015internet}. A visual representation of the generic three-layer architecture is featured in Fig. \ref{fig:arch_three_layer}. The details are described below.

\begin{enumerate}
    \item \textit{Perception Layer}: The perception layer, also known as the sensor layer, is the foundational layer of IoT architecture \cite{atzori2012social}. This layer engages with smart devices, including but not limited to smartwatches and smart rings, employing an array of sensors and actuators. The principal objective of this layer pertains to the collection of data from these intelligent devices via sensors, subsequently transmitting the acquired data to the upper layer known as the network layer.
    
    \item \textit{Network Layer}: The network layer, also known as the transmission layer, is the middle layer of the IoT architecture \cite{leo2014federated}. This layer is responsible for receiving the information passed from the perception layer and determining the routes to transmit the processed data to various connected IoT devices and applications using integrated networks such as wired or wireless secure connections. The network layer is the core layer of the IoT three-layer architecture, as it uses various devices such as routing devices, gateways, switches, and hubs and operates them by using various communication technologies such as WiFi, Bluetooth, 3G, LTE, Zigbee, etc. In summary, the network layer is responsible for transmitting data to and from several applications through interfaces and gateways using multiple communication technologies and protocols. 
    \item \textit{Application Layer}: Serving as the uppermost tier within the IoT architecture, denoted as the application layer or business layer, as referenced in \cite{al2015}, this layer is tasked with the aggregation of data from the network layer, thereby striving to attain the objective of establishing a smart environment, the ultimate aim of the IoT paradigm. 
    This layer accommodates a diverse array of applications, each 
    characterized by their own requirements. Examples of such applications are smart grids, smart cities, and smart transportation, as elaborated in \cite{mahmoud2015internet}. Moreover, this layer assumes the responsibility of upholding the data's authenticity, integrity, and confidentiality, as elaborated in \cite{wu2010research}.
\end{enumerate}
\begin{figure}[t]
        \centering
        \includegraphics[width=0.7\textwidth]{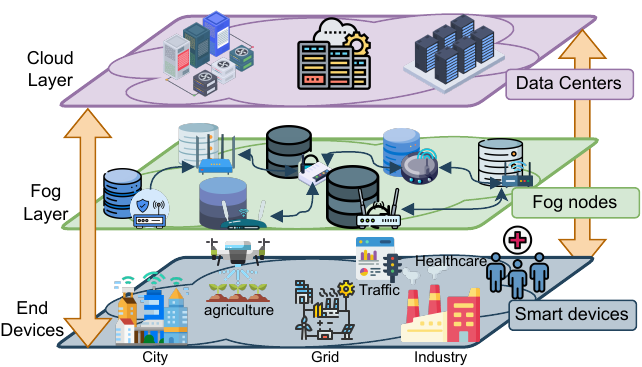}
        \caption{Visual representation of Fog vs Cloud}
        \label{fig:fogVsCloud}
    \end{figure}

\noindent The three-layer architecture is the generalized and most common architecture, and several systems have integrated this architecture \cite{wu2010research}. Although this multi-layer architecture seems simple at first glance, the functionalities of the network and application layers might get complex at times. For example, the network layer is not only responsible for data transmission but also provides data services such as data aggregation and processing, etc. On the other hand, the application layer is not solely responsible for providing service to customers and users but also provides data analysis, conducts data mining, etc. Therefore, in response to specific requirements, additional layers have been incorporated, building upon the fundamental layers. For instance, the four-layered or five-layered architectures enhance the system's flexibility. Nonetheless, the three-layer architecture serves as the foundation for all these variations. Furthermore, new-generation applications require shorter response times and low energy consumption as IoT devices have limited capacity \cite{shi2016edge,nour2020compute}. Therefore, researchers have utilized fog and cloud layers, which are visualized in Fig. \ref{fig:fogVsCloud}.

\section{Key Technologies of IoT}
\label{sec:technologies}
\begin{figure}[ht]
    \centering
    \includegraphics[width=0.85\textwidth]{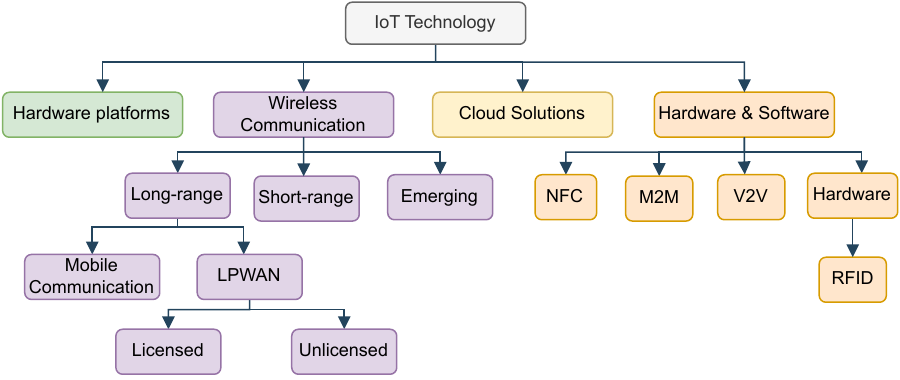}
    \caption{Technology Taxonomy of IoT}
    \label{fig:technology}
\end{figure}

IoT requires a variety of technologies to operate, which are deployed in various layers of an IoT architecture, and there are many different sorts of technologies, including hardware technology, software technology, and, most crucially, communication technology. 
This section discusses crucial IoT technologies that are used to ensure the successful operation of an IoT system. Fig. \ref{fig:technology} depicts the taxonomy of IoT technologies.


\subsection{Hardware Platforms}
The major components of an IoT system are devices attached to sensors or wearable devices that are used for data collection. Therefore, various types of hardware platforms are used in order to build these sensor devices. Several key points need to be considered before selecting the hardware platforms, such as the purpose of the IoT device or the type of connectivity it requires.

The two most accessible and popular hardware platforms are the Raspberry Pi and Arduino. Both of them have strong data acquisition, processing, and storage capabilities and provide both wireless and wired connectivity. However, in terms of power management, Arduino is superior to Raspberry Pi because Raspberry Pi does not contain sleep or suspend modes for power utilization, while Arduino does \cite{maggie2021best}. Intel Galileo Gen and Intel Edison are examples of using an Arduino IDE.


\subsection{Wireless Communication Technology}
Given the vast number of devices in the current IoT network and the expectation of diverse connected devices in future IoT applications, there's an urge to develop various technologies to facilitate their connectivity. This subsection explores the existing wireless technologies designed for IoT connectivity and categorizes them into three groups.

\begin{table}[t]
\caption{Comparison between popular short-range technologies based on various parameters.}
\label{tab:comparisonSRT}
\resizebox{\textwidth}{!}{%
\begin{tabular}{l|llllll}
\hline
\rowcolor[gray]{.9} 
\multicolumn{1}{c}{\cellcolor[gray]{.9}} &
  \multicolumn{1}{c}{\cellcolor[gray]{.9}} &
  \multicolumn{1}{c}{\cellcolor[gray]{.9}} &
  \multicolumn{1}{c}{\cellcolor[gray]{.9}} &
  \multicolumn{1}{c|}{\cellcolor[gray]{.9}} &
  \multicolumn{2}{c}{\cellcolor[gray]{.9}\textbf{OWC}} \\ \cline{6-7} 
\rowcolor[gray]{.9} 
\multicolumn{1}{c}{\multirow{-2}{*}{\cellcolor[gray]{.9}\textbf{Parameters}}} &
  \multicolumn{1}{c}{\multirow{-2}{*}{\cellcolor[gray]{.9}\textbf{Bluetooth}}} &
  \multicolumn{1}{c}{\multirow{-2}{*}{\cellcolor[gray]{.9}\textbf{ZigBee}}} &
  \multicolumn{1}{c}{\multirow{-2}{*}{\cellcolor[gray]{.9}\textbf{LR-WPAN}}} &
  \multicolumn{1}{c|}{\multirow{-2}{*}{\cellcolor[gray]{.9}\textbf{Wi-Fi}}} &
  \multicolumn{1}{c|}{\cellcolor[gray]{.9}\textbf{VLC}} &
  \multicolumn{1}{c}{\cellcolor[gray]{.9}\textbf{BS-ILC}} \\ \hline \hline
Standard &
  IEEE 802.15.1 &
  IEEE 802.15.4 &
  IEEE 802.15.4 &
  \begin{tabular}[c]{@{}l@{}}IEEE 802.11a/b/c/\\ d/g/n/ac/ah\end{tabular} &
  \multicolumn{1}{l}{\begin{tabular}[c]{@{}l@{}}IEEE 802.15.7m\\ IEEE 802.15.13\end{tabular}} &
  LoRaWAN \\ 
\rowcolor[gray]{.96}Frequency Band &
  1MHz - 2.48GHz &
  \begin{tabular}[c]{@{}l@{}}Mainly at 2.4 GHz\\ Optionally 868MHz\\  or 915MHz\end{tabular} &
  \begin{tabular}[c]{@{}l@{}}868/915 MHz,\\ 2.4 GHz\end{tabular} &
  \begin{tabular}[c]{@{}l@{}}a: 5GHz, b: 2.4GHz,\\ g: 2.4GHz, n: 2.4GHz\\ 802.11ah: 1/2/16MHz\end{tabular} &
  \multicolumn{1}{l}{400-800 THz} &
  \begin{tabular}[c]{@{}l@{}}Varies by region, \\ Europe: 868 MHz\\ USA: 915 MHz\end{tabular} \\ 
Data Rate &
  1Mbps - 3 Mbps &
  20kbps to 250kbps &
  40-250 kbps &
  \begin{tabular}[c]{@{}l@{}}a: 54Mbps, b: 11Mbps,\\ g: 54Mbps, ah: 300Mbps\\ n: 600Mbps, ac: 7Gbps\end{tabular} &
  \multicolumn{1}{l}{\begin{tabular}[c]{@{}l@{}}15.13: multi gigabit\\ Recent: 100 Gbps\end{tabular}} &
  100Gbps \\ 
\rowcolor[gray]{.96}Transmission range &
  \begin{tabular}[c]{@{}l@{}}Classic: 100m\\ BLE: 240m\end{tabular} &
  10-100 meters &
  10-100 meters &
  100m to 1km &
  \multicolumn{1}{l}{\begin{tabular}[c]{@{}l@{}}Typically, within a room\\ 7m: 200 meters\\ 15.13: Several meters\end{tabular}} &
  200 meters \\ 
Energy consumption &
  \begin{tabular}[c]{@{}l@{}}Classic: High\\ BLE: Low\end{tabular} &
  Low &
  Low &
  Moderate to high &
  \multicolumn{1}{l}{\begin{tabular}[c]{@{}l@{}}Transmitters: Moderate\\ Receivers:Minimal\end{tabular}} &
  Very low \\ 
\rowcolor[gray]{.96}Cost &
  Cost-effective &
  cost-effective &
  Cost-effective &
  moderate to high &
  \multicolumn{1}{l}{moderate to high} &
  Cost-effective \\ 
RA protocol &
  \begin{tabular}[c]{@{}l@{}}FHSS, FDMA\\ TDMA based polling\end{tabular} &
  CSMA/CA &
  CSMA/CA &
  CSMA/CD, CSMA/CA &
  \multicolumn{1}{l}{\begin{tabular}[c]{@{}l@{}}CSMA/CA, \\ TDMA/CDMA\end{tabular}} &
  LBT \\ 
\rowcolor[gray]{.96}Modulation type &
  \begin{tabular}[c]{@{}l@{}}GFSK, DQPSK, \\ $\pi/4$-DQPSK\end{tabular} &
  BPSK/O-QPSK &
  BPSK/O-QPSK &
  BPSK/QPSK/QAM &
  \multicolumn{1}{l}{OOK/PPM/OFDM} &
  CSS \\ \hline
\end{tabular}%
}
\end{table}

\vspace{0.5cm}
\noindent\textit{Short-Range Technologies}\\
Short-range technologies are commonly used in IoT applications to enable communication between devices within a limited proximity. These technologies are well-suited for scenarios where devices need to exchange data within a small coverage area, typically spanning from a few meters to a few kilometers. There are several short-range technologies, each with unique characteristics and advantages suited for specific purposes. Table \ref{tab:comparisonSRT} provides a comprehensive overview of various short-range technologies used in the IoT environment, categorized based on various parameters such as frequency band, data rate, transmission range, and more. The table summarizes the technical specifications of Bluetooth, ZigBee, Wi-Fi, LR-WPAN, VLC, and BS-ILC \cite{ding2020iot, ray2018survey}.
\begin{landscape}
\begin{table}[]
\centering
\caption{Comparison between popular long-range technologies based on various parameters.}
\label{tab:longRangeTech}
\resizebox{\columnwidth}{!}{%
\begin{tabular}{l|lllll|llll}
\hline
\rowcolor[gray]{0.9} 
\multicolumn{1}{c|}{\cellcolor[gray]{0.9}} &
  \multicolumn{5}{c|}{\cellcolor[gray]{0.9}\textbf{Mobile Communication Technology}} &
  \multicolumn{4}{c}{\cellcolor[gray]{0.9}\textbf{LPWAN TECHNOLOGIES}} \\ \cline{2-10} 
\rowcolor[gray]{0.9} 
\multicolumn{1}{c|}{\cellcolor[gray]{0.9}} &
  \multicolumn{1}{c}{\cellcolor[gray]{0.9}} &
  \multicolumn{1}{c}{\cellcolor[gray]{0.9}} &
  \multicolumn{1}{c}{\cellcolor[gray]{0.9}} &
  \multicolumn{1}{c}{\cellcolor[gray]{0.9}} &
  \multicolumn{1}{c|}{\cellcolor[gray]{0.9}} &
  \multicolumn{2}{c|}{\cellcolor[gray]{0.9}\textbf{UNLICENSED LPWAN}} &
  \multicolumn{2}{c}{\cellcolor[gray]{0.9}\textbf{LICENSED LPWAN}} \\ \cline{7-10} 
\rowcolor[gray]{0.9}
\multicolumn{1}{c|}{\multirow{-3}{*}{\cellcolor[gray]{0.9}\textbf{\begin{tabular}[c]{@{}c@{}}\diagbox{Parameters}{Technologies}\end{tabular}}}} &
  \multicolumn{1}{c}{\multirow{-2}{*}{\cellcolor[gray]{0.9}\textbf{2G}}} &
  \multicolumn{1}{c}{\multirow{-2}{*}{\cellcolor[gray]{0.9}\textbf{3G}}} &
  \multicolumn{1}{c}{\multirow{-2}{*}{\cellcolor[gray]{0.9}\textbf{4G}}} &
  \multicolumn{1}{c}{\multirow{-2}{*}{\cellcolor[gray]{0.9}\textbf{5G}}} &
  \multicolumn{1}{c|}{\multirow{-2}{*}{\cellcolor[gray]{0.9}\textbf{WiMax}}} &
  \multicolumn{1}{c|}{\cellcolor[gray]{0.9}\textbf{LoRa}} &
  \multicolumn{1}{c|}{\cellcolor[gray]{0.9}\textbf{Sigfox}} &
  \multicolumn{1}{c|}{\cellcolor[gray]{0.9}\textbf{LTE-M}} &
  \multicolumn{1}{c}{\cellcolor[gray]{0.9}\textbf{NB-IoT}} \\ \hline \hline
Standard &
  \multicolumn{1}{l}{GSM, CDMA} &
  \multicolumn{1}{l}{\begin{tabular}[c]{@{}l@{}}UMTS, \\ CDMA2000\end{tabular}} &
  \multicolumn{1}{l}{\begin{tabular}[c]{@{}l@{}}LTE, LTE-A, \\ IEEE 802.16\end{tabular}} &
  \multicolumn{1}{l}{\begin{tabular}[c]{@{}l@{}}5G NR, \\ Wi-Fi 6 \\ (802.11ax)\end{tabular}} &
  IEEE 802.16 &
  \multicolumn{1}{l}{LoRaWAN R1.0} &
  \multicolumn{1}{l|}{\begin{tabular}[c]{@{}l@{}}Proprietary \\ technology\end{tabular}} &
  \multicolumn{1}{l}{3GPP LTE} &
  3GPP LTE \\ 
\rowcolor[gray]{0.96} 
\begin{tabular}[c]{@{}l@{}}Frequency band\end{tabular} &
  \multicolumn{1}{l}{\cellcolor[gray]{0.96}\begin{tabular}[c]{@{}l@{}}850MHz, \\ 900MHz, \\ 1800MHz, \\ 1900MHz\end{tabular}} &
  \multicolumn{1}{l}{\cellcolor[gray]{0.96}\begin{tabular}[c]{@{}l@{}}850MHz, \\ 1900MHz, \\ 2100MHz\end{tabular}} &
  \multicolumn{1}{l}{\cellcolor[gray]{0.96}\begin{tabular}[c]{@{}l@{}}700MHz, \\ 1700/2100MHz,\\ 2500MHz\end{tabular}} &
  \multicolumn{1}{l}{\cellcolor[gray]{0.96}\begin{tabular}[c]{@{}l@{}}60MHz - \\ 80GHz\end{tabular}} &
  \begin{tabular}[c]{@{}l@{}}2.3GHz, \\ 2.5GHz, \\ 3.5GHz\end{tabular} &
  \multicolumn{1}{l}{\cellcolor[gray]{0.96}\begin{tabular}[c]{@{}l@{}}Unlicensed ISM\\ bands, 125kHz, \\ 250kHz\end{tabular}} &
  \multicolumn{1}{l|}{\cellcolor[gray]{0.96}\begin{tabular}[c]{@{}l@{}}Unlicensed ISM \\ bands, 100Hz\end{tabular}} &
  \multicolumn{1}{l}{\cellcolor[gray]{0.96}\begin{tabular}[c]{@{}l@{}}Licensed LTE\\ bands, 1.4MHz\end{tabular}} &
  \begin{tabular}[c]{@{}l@{}}Licensed LTE\\ bands, 200kHz\end{tabular} \\ 
Data rate &
  \multicolumn{1}{l}{\begin{tabular}[c]{@{}l@{}}9Kbps - \\ 384Kbps\end{tabular}} &
  \multicolumn{1}{l}{\begin{tabular}[c]{@{}l@{}}384Kbps - \\ several Mbps\end{tabular}} &
  \multicolumn{1}{l}{\begin{tabular}[c]{@{}l@{}}100Mbps - \\ 1Gbps\end{tabular}} &
  \multicolumn{1}{l}{\begin{tabular}[c]{@{}l@{}}1Gbps - \\ 20Gbps or \\ higher\end{tabular}} &
  \begin{tabular}[c]{@{}l@{}}16d: 75Mbps\\ 16e: 1Gbps\end{tabular} &
  \multicolumn{1}{l}{\begin{tabular}[c]{@{}l@{}}0.3Kbps - \\ 50Kbps\end{tabular}} &
  \multicolumn{1}{l|}{\begin{tabular}[c]{@{}l@{}}100bps - \\ 600bps\end{tabular}} &
  \multicolumn{1}{l}{\begin{tabular}[c]{@{}l@{}}300bps - \\ 1Mbps\end{tabular}} &
  \begin{tabular}[c]{@{}l@{}}200bps - \\ 250kbps\end{tabular} \\ 
\rowcolor[gray]{0.96} 
\begin{tabular}[c]{@{}l@{}}Transmission range\end{tabular} &
  \multicolumn{1}{l}{\cellcolor[gray]{0.96}\begin{tabular}[c]{@{}l@{}}several \\ kilometers\end{tabular}} &
  \multicolumn{1}{l}{\cellcolor[gray]{0.96}\begin{tabular}[c]{@{}l@{}}several \\ kilometers\end{tabular}} &
  \multicolumn{1}{l}{\cellcolor[gray]{0.96}\begin{tabular}[c]{@{}l@{}}several \\ kilometers\end{tabular}} &
  \multicolumn{1}{l}{\cellcolor[gray]{0.96}\begin{tabular}[c]{@{}l@{}}several \\ kilometers\end{tabular}} &
  \begin{tabular}[c]{@{}l@{}}several \\ kilometers\end{tabular} &
  \multicolumn{1}{l}{\cellcolor[gray]{0.96}\begin{tabular}[c]{@{}l@{}}Urban: 5km\\ Rural: 20km\end{tabular}} &
  \multicolumn{1}{l|}{\cellcolor[gray]{0.96}\begin{tabular}[c]{@{}l@{}}Urban: 10km\\ Rural: 50km\end{tabular}} &
  \multicolumn{1}{l}{\cellcolor[gray]{0.96}\begin{tabular}[c]{@{}l@{}}Urban: 1km\\ Rural: 10km\end{tabular}} &
  \begin{tabular}[c]{@{}l@{}}several \\ kilometers\end{tabular} \\ 
\begin{tabular}[c]{@{}l@{}}Energy consumtion\end{tabular} &
  \multicolumn{1}{l}{low} &
  \multicolumn{1}{l}{Moderate} &
  \multicolumn{1}{l}{\begin{tabular}[c]{@{}l@{}}Moderate \\ to high\end{tabular}} &
  \multicolumn{1}{l}{\begin{tabular}[c]{@{}l@{}}Low (Energy-\\ efficient)\end{tabular}} &
  high &
  \multicolumn{1}{l}{Extremely low} &
  \multicolumn{1}{l|}{Low} &
  \multicolumn{1}{l}{Moderate} &
  Low \\ 
\rowcolor[gray]{0.96} 
RA protocol &
  \multicolumn{1}{l}{\cellcolor[gray]{0.96}\begin{tabular}[c]{@{}l@{}}TDMA/\\ CDMA/\\ FDMA\end{tabular}} &
  \multicolumn{1}{l}{\cellcolor[gray]{0.96}\begin{tabular}[c]{@{}l@{}}CDMA/\\ WCDMA\end{tabular}} &
  \multicolumn{1}{l}{\cellcolor[gray]{0.96}\begin{tabular}[c]{@{}l@{}}OFDMA/\\ SC-FDMA\end{tabular}} &
  \multicolumn{1}{l}{\cellcolor[gray]{0.96}\begin{tabular}[c]{@{}l@{}}Massive MIMO/\\ beamforming/\\ CP-OFDM\end{tabular}} &
  \begin{tabular}[c]{@{}l@{}}OFDMA/\\ TDD/FDD\end{tabular} &
  \multicolumn{1}{l}{\cellcolor[gray]{0.96}\begin{tabular}[c]{@{}l@{}}ALOHA/\\ Slotted-ALOHA\end{tabular}} &
  \multicolumn{1}{l|}{\cellcolor[gray]{0.96}ALOHA} &
  \multicolumn{1}{l}{\cellcolor[gray]{0.96}Slotted-ALOHA} &
  Slotted-ALOHA \\ 
Modulation type &
  \multicolumn{1}{l}{GMSK/QPSK} &
  \multicolumn{1}{l}{QPSK/16-QAM} &
  \multicolumn{1}{l}{\begin{tabular}[c]{@{}l@{}}OFDM/QPSK/\\ 16-QAM/64-QAM/\\ 256-QAM\end{tabular}} &
  \multicolumn{1}{l}{256-QAM} &
  \begin{tabular}[c]{@{}l@{}}OFDM, QPSK, \\ 16-QAM, \\ 64-QAM\end{tabular} &
  \multicolumn{1}{l}{CSS} &
  \multicolumn{1}{l|}{GFSK/DBPSK} &
  \multicolumn{1}{l}{\begin{tabular}[c]{@{}l@{}}QPSK/QAM/\\ BPSK\end{tabular}} &
  QPSK/BPSK \\ 
\rowcolor[gray]{0.96} 
Cost &
  \multicolumn{1}{l}{\cellcolor[gray]{0.96}Cost-effective} &
  \multicolumn{1}{l}{\cellcolor[gray]{0.96}Moderate} &
  \multicolumn{1}{l}{\cellcolor[gray]{0.96}high} &
  \multicolumn{1}{l}{\cellcolor[gray]{0.96}high} &
  high &
  \multicolumn{1}{l}{\cellcolor[gray]{0.96}Cost-effective} &
  \multicolumn{1}{l|}{\cellcolor[gray]{0.96}\begin{tabular}[c]{@{}l@{}}Depends on \\ subscription\\ model\end{tabular}} &
  \multicolumn{1}{l}{\cellcolor[gray]{0.96}Moderate to high} &
  Moderate to high \\ \hline
\end{tabular}%
}
\end{table}
\end{landscape}

Bluetooth, ZigBee, LR-WPAN, and BS-ILC offer cost-effective communication solutions, whereas the overall cost of Wi-Fi and VLC may vary depending on usage. In terms of frequency bands, most technologies operate around 2.4 GHz, except for VLC, which utilizes a much larger frequency band. The frequency of BS-ILC varies by region. VLC and BS-ILC prioritize achieving high transmission rates of approximately 100 Gbps, while Wi-Fi transmission rates can vary across different standards. Notably, VLC and BS-ILC exhibit very low energy consumption, as indicated in the table.

\vspace{0.5cm}
\noindent\textit{Long Range Technologies}\\
Long-range technologies are designed to transmit data or signals over considerable distances, typically much farther than short-range technologies like Bluetooth or Wi-Fi. Long-range technologies are commonly used in various applications, such as long-distance wireless communication, remote monitoring, tracking, and more, and are essential for scenarios where data needs to be transmitted reliably over extensive geographical areas. These technologies can be classified into two main categories: mobile communication technology and LPWAN technologies. Table \ref{tab:longRangeTech} presents a detailed overview of various long-range technologies utilized in the IoT environment. It provides a summary of the technical specifications for various versions of mobile communication technologies as well as different versions of LPWAN technologies.

It is evident from the table that when it comes to mobile communication technologies, 2G, 3G, and 4G operate within a frequency band ranging from approximately 850MHz to 2100MHz, while 5G utilizes a much broader frequency band of up to 80GHz. Additionally, of all five technological variations, 5G offers the highest data rate, exceeding 20Gbps. In terms of energy efficiency, both 2G and 5G have low consumption rates, whereas WiMax exhibits high energy consumption. However, it is worth noting that 3G, 4G, and 5G all come with higher implementation costs. In the UNLICENSED LPWAN category, the top contenders are LoRa and Sigfox, while the most promising options in the LICENSED LPWAN technology are LTE-M (Long-Term Evolution for Machines) and NB-IoT (Narrowband IoT). The key distinction highlighted in the table is that UNLICENSED LPWAN technology operates within unlicensed spectrum resources, leading to lower deployment costs, whereas LICENSED LPWAN technology relies on licensed spectrum resources, resulting in relatively higher expenses for both devices and deployment \cite{ding2020iot}.

\vspace{0.5cm}
\noindent\textit{Emerging For Massive Connectivity}\\
    While current wireless IoT technologies have achieved some success in supporting various IoT applications, they still face challenges in meeting the future demands of IoT. For example, handling the connectivity among a vast number of IoT devices with a limited transmission payload and several random access protocols used by existing technologies often results in significant problems such as frequent access collisions, increased delays, and a high amount of signaling overhead for IoT devices. Furthermore, the constrained availability of wireless resources for connecting IoT devices creates shortages and inefficient utilization of these resources. Hence, there have been numerous ongoing initiatives to tackle the limitations of current technologies. Among these, the solutions CS, NOMA, mMIMO, and ML stand out as the most promising \cite{ding2020iot}. 

    These emerging technologies offer the capability not only to support massive connectivity but also to deliver high reliability and low-latency transmissions. However, it is essential to acknowledge that there are existing issues and constraints that must be resolved for their effective implementation. It is anticipated that the development of even more advanced technologies will address critical IoT challenges. Simultaneously, efforts should focus on intelligently integrating existing and emerging technologies to unlock their full potential and optimize system performance \cite{ding2020iot}.


\subsection{Cloud Solutions}
IoT cloud solutions offer various services like real-time data collection, data transmission, monitoring, data analytics, improved decision-making, and device management. These services are available on a pay-as-you-use basis, allowing users to pay only for the services they actually use. Cloud platforms can be integrated into numerous domains, including healthcare, smart cities, agriculture, education, and supply chain management. For simplicity, Table \ref{tab:cloudPlatform} presents several popular platforms in the agriculture domain \cite{ray2018survey}.

\begin{table}[ht]
\centering
\caption{Popular IoT cloud solutions in agriculture domain}
\label{tab:cloudPlatform}
\resizebox{\textwidth}{!}{%
\begin{tabular}{l|lccccl}
\hline
\rowcolor[gray]{.9}\multicolumn{1}{c|}{\textbf{\begin{tabular}[c]{@{}c@{}}IoT Cloud \\ platform\end{tabular}}} &
  \multicolumn{1}{c}{\textbf{Provider}} &
  \textbf{\begin{tabular}[c]{@{}c@{}}Real-time \\ Data Capture\end{tabular}} &
  \textbf{\begin{tabular}[c]{@{}c@{}}Data \\ Visualization\end{tabular}} &
  \textbf{\begin{tabular}[c]{@{}c@{}}Cloud \\ Service Type\end{tabular}} &
  \textbf{\begin{tabular}[c]{@{}c@{}}Data \\ Analytics\end{tabular}} &
  \multicolumn{1}{c}{\textbf{\begin{tabular}[c]{@{}c@{}}Developer\\ Cost\end{tabular}}} \\ \hline \hline
ThingSpeak &
  MathWorks &
  Yes &
  Yes (Matlab) &
  Public &
  Yes &
  Free \\ 
\rowcolor[gray]{.96}Plotly &
  \begin{tabular}[c]{@{}l@{}}Plotly \\ Technologies Inc.\end{tabular} &
  Yes &
  Yes &
  Public &
  Yes &
  Free \\ 
Carriots &
  Altair &
  Yes &
  Yes &
  Private &
  No &
  \begin{tabular}[c]{@{}l@{}}Limited up to:\\  10 devices\end{tabular} \\ 
\rowcolor[gray]{.96}Exosite &
  Exosite &
  Yes &
  Yes &
  IoTSaaS &
  Yes &
  2 devices \\ 
GroveStreams &
  \begin{tabular}[c]{@{}l@{}}GroveStreams\\  LLC\end{tabular} &
  Yes &
  Yes &
  Private &
  Yes &
  \begin{tabular}[c]{@{}l@{}}Limited up to:\\ 20 stream, \\ 10K transaction,\\ 5 SMS, 500 Email\end{tabular} \\ 
\rowcolor[gray]{.96}ThingWorx &
  PTC &
  Yes &
  Yes &
  Private &
  Yes &
  Pay per use \\ 
Nimbits &
  Open-source &
  Yes &
  Yes &
  Hybrid &
  No &
  Free \\ \hline
\end{tabular}%
}
\end{table}

\subsection{Hardware and Software Technology}
This category encompasses a range of hardware and software technologies responsible for different aspects of IoT communication channels. Several technologies, such as NFC and M2M, are considered both hardware and software technologies, whereas RFID is classified as a hardware-only technology. The subsequent sections will provide brief insights into these technologies.

\vspace{0.5cm}
\noindent\textit{RFID}\\
 RFID is a technology that consists of one or more readers and several RFID tags. These tags are small microchips, have unique codes, and can be on items like products in a store or even access cards. When an RFID reader device sends out electromagnetic radio waves, the tags respond with their unique Electronic Product Codes (EPCs) \cite{shah2016survey}. The reader captures these codes and sends them to a computer, which can then figure out what the tags belong to. This technology is used in various ways, like keeping track of inventory in stores, allowing access to secure areas, or monitoring packages as they move through the delivery process.
 
 

\vspace{0.5cm}
\noindent\textit{Near Field Communication (NFC)}\\
NFC, an extension of RFID technology, is a short-range wireless connectivity technology that uses magnetic field induction to establish communication between devices when they are brought within close proximity of each other without the need for prior connection establishment. NFC chips are equipped in most modern phones, supporting applications like Apple Pay and Google Pay. NFC operates in the unlicensed radio frequency band at 13.56MHz. Its typical range is about 20 meters, with the actual distance often determined by the size of the antenna in the device. It is expected that NFC technology will play a vital role in the future of IoT by providing a wireless connection tool to link with other smart objects. For instance, using NFC on a mobile device, a user can transform their phone into various objects, such as using it as a credit card for transactions \cite{shah2016survey}.

\vspace{0.5cm}
\noindent\textit{M2M}\\
M2M communication is gaining popularity and involves direct communication between computers, embedded processors, smart sensors, actuators, and mobile devices. It comprises four main components: \textit{sensing}, \textit{heterogeneous access}, \textit{information processing}, and \textit{applications and processing}. In practical terms, M2M functions within a five-part framework: a device for responding to requests, gateways to interact and connect, an area network for providing connectivity between devices and gateways, applications serving as middleware, and a communication network to facilitate communication between gateways and applications. M2M technology is applied across various sectors, including healthcare, smart robotics, cyber transportation systems (CTS), manufacturing systems, smart home technologies, and smart grids \cite{shah2016survey}.





\vspace{0.5cm}
\noindent\textit{Vehicle-to-Vehicle Communications (V2V)}\\
V2V Communications treats each vehicle as a node and enables wireless data exchange (omnidirectional) between vehicles regarding their speed, location, and more. Vehicles equipped with appropriate software, often referred to as safety applications, can use messages from nearby vehicles to detect potential collision risks in real-time. There are two types of communication within this network: one is between vehicles, and the other involves road infrastructure. However, the structure or arrangement of this communication is flexible as vehicles move from one location to another. This network can be divided into four main categories: \textit{safety and collision avoidance}, \textit{traffic infrastructure management}, \textit{vehicle telematics}, and \textit{entertainment services with Internet connectivity} \cite{shah2016survey}.

\section{Applications of IoT}
\label{sec:applications}
IoT has been effectively integrated into various domains, leading to the development of intelligent applications. While some of these applications are already available, others are still in the research phase. Nevertheless, the essence of these applications indicates that IoT is set to enhance people's lives by providing convenience and flexibility. The following section has categorized and provided brief descriptions of the areas where IoT integration has been implemented or proposed.

\subsection{Smart Cities}
The concept of smart cities can be seen as a complex IoT paradigm where the management of public affairs incorporates the introduction of information and communication technology (ICT) solutions \cite{zanella2014internet}. A smart city utilizes public resources to digitize the city and enhance the quality of life. A real-world example of a smart city implementation is ``Padova Smart City'', which has been put into practice in the city of Padova, Italy \cite{cenedese2014padova}. Barcelona and Stockholm are two noteworthy examples of smart cities. Barcelona has embarked on the CityOS project, with the primary goal of creating a centralized operating system to manage all the smart devices and services available within the city. Their focus has predominantly been on enhancing smart transportation and water systems. Similarly, Stockholm has also placed significant emphasis on these two domains and holds the distinction of being one of the pioneering cities to implement the concept of congestion charges, wherein users are imposed fees for entering congested areas \cite{sethi2017internet}. In addition, Table \ref{tab:smartCities} summarizes related research works in this domain.


\begin{table}[!ht]
\centering
\caption{Relevant applications in different Smart Cities domain}
\label{tab:smartCities}
\resizebox{\columnwidth}{!}{%
\begin{tabular}{l|l||l|l}
\hline
\rowcolor[gray]{.9}\textbf{Applications} & \textbf{Related Works} & \textbf{Applications} & \textbf{Related Works} \\ \hline \hline
Structural health of buildings & \cite{lynch2006summary, zanella2014internet} & Waste management & \cite{zanella2014internet} \\ 
\rowcolor[gray]{.96}\begin{tabular}[c]{@{}l@{}}Air quality and noise \\ monitoring\end{tabular} & \cite{maisonneuve2009citizen, li2008performance} & \begin{tabular}[c]{@{}l@{}}Smart Transport and \\ Trafﬁc congestion\end{tabular} & \cite{sethi2017internet, hsieh2006automatic, hu2014mood, kotb2017smart, mimbela2007summary} \\ 
Smart grid & \cite{al2019iot, lin2015towards} & Smart water systems & \cite{hauber2006smart, sethi2017internet} \\ \hline
\end{tabular}%
}
\end{table}


\noindent\textit{Structural Health of Buildings}\\
This service involves continuous monitoring of areas prone to various external factors and measuring the health of any building. IoT sensors connected to the buildings can store information about the building's strength, which will help analyze how sturdy the building is or if it requires any refinement \cite{lynch2006summary}. Depending on the usage, one can employ various kinds of sensors, including those for tracking vibrations to measure the stress on the building, temperature and humidity sensors, or other types of atmospheric sensors to assess the level of pollution in an area \cite{zanella2014internet}. Employing IoT in this area can reduce manual labor, where humans have to manually assess the building's health and environmental conditions, and it can also reduce the overall cost to a great extent.

\vspace{0.5cm}
\noindent\textit{Waste Management}\\
Waste management is a crucial operation in any city, whether it is considered a smart city or not, as it directly impacts the livability of the area. Thus, an efficient waste management system is essential for any society. Integrating the IoT into waste management offers multiple benefits. It enables the detection of waste levels and real-time tracking of garbage truck routes, leading to more efficient route planning. Moreover, it can streamline the manual labor involved in waste separation and monitor the disposal process.  Sensors on garbage vehicles connected to a central software system achieve these tasks by analyzing and controlling the system based on the data collected \cite{zanella2014internet}. This approach reduces the cost of manual processes and enhances recycling management.

\vspace{0.5cm}
\noindent\textit{Air Quality and Noise Monitoring}\\
Creating a healthy and safe environment for all living beings is crucial, and monitoring air quality and noise levels in any area is a key part of achieving this goal. Various environmental sensors, including soil sensors, temperature and humidity sensors, and gas sensors, can detect the presence of toxic and pollutant substances in the air and assess pollution levels. This information enables local authorities to control pollution, implement effective measures to reduce pollutant levels, identify highly polluted or toxic areas, and designate suitable locations for outdoor activities with good air quality \cite{maisonneuve2009citizen, li2008performance}. Likewise, it is essential to maintain a balanced noise level for all living beings in society, including humans and animals. Using noise sensors to measure decibel levels, the central authority can collect data to identify noisy areas and regulate noise levels to keep them within acceptable limits.

\vspace{0.5cm}
\noindent\textit{Smart Transport and Trafﬁc Congestion}\\
In today's world, traffic congestion is a widespread issue affecting nearly every country and city, particularly as more than half of the world's population now resides in urban areas \cite{sethi2017internet}. To tackle this issue, many cities have adopted IoT solutions to establish smart transportation systems aimed at managing traffic congestion. These initiatives include \textit{smart traffic lights} \cite{sethi2017internet} and \textit{smart parking system} \cite{kotb2017smart, mimbela2007summary}, which collectively enhance transportation capacity and improve safety and speed for travelers. The primary advantages of smart transportation systems are reducing traffic congestion, ensuring hassle-free travel, and facilitating easy parking. Moreover, these systems enable quicker responses in case of accidents and contribute to accident reduction by effectively managing traffic flow \cite{sethi2017internet}. These objectives are achieved through the use of a variety of sensor technologies, including accelerometers for measuring speed, RFIDs for vehicle identification, GPS sensors for location tracking, gyroscopes for direction detection, and cameras for recording traffic patterns and vehicle movements. Some real-world uses of these sensors can be seen in \textit{applications for managing and monitoring traffic} \cite{sethi2017internet, hsieh2006automatic}, \textit{applications to ensure safety} \cite{hu2014mood}, and {\textit{application for detecting accidents}.

 \vspace{0.5cm}
\noindent\textit{Smart grid}\\
A traditional grid system is an electrical grid that includes transmission lines, transformers, and various communication utilities responsible for delivering electricity from power plants to homes or businesses. One significant limitation of the traditional grid is its one-way communication, which prevents power plants from efficiently responding to increasing power demands from consumers. To address this challenge, the smart grid establishes a two-way communication system between utilities and consumers, which enables more effective management of economic, sustainable, and secure power resources \cite{al2019iot}. Integrating IoT into grid systems enables equipping houses and businesses with smart meters that monitor energy generation, storage, and consumption, and transmit this data to the smart grid. 


\vspace{0.5cm}
\noindent\textit{Smart Water Systems}\\
Water is one of the most critical natural resources, and its scarcity is a prevalent issue in many parts of the world. Therefore, implementing smart water systems is not a luxury but a necessity. The primary role of smart water systems is to monitor, measure, and efficiently distribute water usage. Hauber-Davidson and Idris have designed a notable model in this field, the smart water metre \cite{hauber2006smart}. These metres can detect water inflow and outflow and identify any potential leaks. Additionally, smart water metres can utilize data from smart river sensors and weather information to assist in flood prediction \cite{sethi2017internet}.

\begin{table}[b]
\centering
\caption{Relevant works in medical and healthcare applications}
\label{tab:medicalHealthcare}
\resizebox{\columnwidth}{!}{%
\begin{tabular}{l|l||l|l}
\hline
\rowcolor[gray]{.9}\textbf{Applications} & \textbf{Related Works} & \textbf{Applications} & \textbf{Related Works} \\ \hline \hline
ECG Monitoring & \cite{wu2019wearable, bansal2019iot} & Glucose level monitoring & \cite{istepanian2011potential, sunny2018optical} \\ 
\rowcolor[gray]{.96}Temperature monitoring & \cite{ota20173d, gunawan2020design} & Blood pressure monitoring & \cite{xin2017novel, pradhan2021iot} \\ 
Oxygen saturation monitoring & \cite{fu2015system, agustine2018heart} & Mood monitoring & \cite{kaur2019smartemodetect, kushawaha2020sentiment} \\ 
\rowcolor[gray]{.96}Medication management & \cite{bharadwaj2017enhancing, karagiannis2020design} & Wheelchair management & \cite{lee2017real, kumar2020design} \\ 
Rehabilitation & \cite{nave2018smart, jiang2020combination} & Fitness & \cite{sundholm2014smart, seshadri2019wearable} \\ 
\rowcolor[gray]{.96}Other notable application & \begin{tabular}[c]{@{}l@{}}\cite{heshmat2018framework, rodrigues2020new, rajan2020fog, pradhan2020medical},\\  \cite{liu2019deep, cecil2018iomt, su2020internet, bhatia2019towards}\end{tabular} &  &  \\ \hline
\end{tabular}%
}
\end{table}

\subsection{Medical and Healthcare}
The healthcare sector has experienced remarkable advancements through the integration of IoT, offering solutions to real-life healthcare challenges and enhancing people's lifestyles. Researchers have proposed various applications in healthcare that utilize wearable sensor devices to monitor patients' health, diagnose diseases, issue emergency alerts, and notify users when necessary. Remote monitoring saves time for both patients and doctors while reducing overall healthcare costs. Furthermore, sharing data collected from these wearable devices with healthcare researchers contributes to the development of safer and more timely healthcare solutions and aids in the discovery of cures and vaccines for emerging diseases. This section explores several IoT applications in healthcare, and a summary of the associated research papers is presented in Table \ref{tab:medicalHealthcare}.


\vspace{0.5cm}
\noindent\textit{Electrocardiogram (ECG) Monitoring}\\
An ECG measures the heart's electrical signals and serves as an indicator of heart health, aiding in the detection of conditions like arrhythmia, prolonged QT interval, and myocardial ischemia. An interesting example would be Wu et al.'s ECG data monitoring system \cite{wu2019wearable}, where a bipotential chip is utilized by attaching to the user's t-shirt and transmitting data to a smartphone via Bluetooth. Combining IoT systems with big data analytics enables real-time ECG data monitoring.



\vspace{0.5cm}
\noindent\textit{Glucose Level Monitoring}\\
Diabetes is a medical condition characterized by higher blood glucose levels than those found in individuals without diabetes. Among various approaches to identifying diabetes, the fingerstick method, involving a small pinprick to the fingertip followed by blood glucose level measurement, remains the most commonly used diagnostic approach. The advent of IoT technology has brought about improvements in this process, making it quicker and more convenient for patients. Istepanian et al. \cite{istepanian2011potential} have introduced an IoT-integrated noninvasive blood glucose monitoring device to continuously monitor glucose levels, eliminating the need for fingerstick testing. It is worth noting that optical sensors, such as infrared LED and near-infrared photodiode setups, have also been used for glucose level measurements. 



\vspace{0.5cm}
\noindent\textit{Temperature Monitoring}\\
Traditionally, temperature measurement methods involve using thermometers placed in the mouth, ear, or rectum, but these methods often cause discomfort for the patient and pose an elevated risk of infection. However, recent advancements in IoT-based temperature monitoring applications have effectively addressed these issues. For instance, authors in \cite{ota20173d} have introduced a 3D-printed wearable device designed to be inserted into the ear. This device utilizes an infrared sensor to measure temperature from the tympanic membrane, ensuring accuracy while remaining environment-independent.

\vspace{0.5cm}
\noindent\textit{Blood Pressure Monitoring}\\
In many diagnostic processes, measuring blood pressure is a compulsory step. However, the major issue with the traditional method is that it requires one person to record the blood pressure. Therefore, the integration of IoT into blood pressure monitoring has been a blessing for both doctors and patients. A wearable cuffless gadget \cite{xin2017novel}, for example, is capable of measuring both systolic and diastolic pressure, with the results stored in the cloud.

\vspace{0.5cm}
\noindent\textit{Oxygen Saturation Monitoring}\\
Pulse oximetry, a highly beneficial noninvasive device for measuring oxygen saturation, addresses the limitations of traditional methods and allows for real-time monitoring. The integration of IoT-based technology has led to significant advancements in pulse oximetry, particularly in the healthcare industry. In a study \cite{fu2015system}, an advanced noninvasive pulse oximetry system has been proposed, capable of measuring oxygen levels, heart rate, and pulse parameters while transmitting this data to a central server.

\vspace{0.5cm}
\noindent\textit{Mood Monitoring}\\
The integration of IoT into the mood monitoring domain offers numerous advantages. IoT can detect a person's mental state by analyzing heartbeats through wearable devices. Kaur et al. \cite{kaur2019smartemodetect} have proposed a wearable device capable of tracking a driver's emotions, including anger, stress, terror, and sadness. The intelligent system, by analyzing emotion variations, determines whether the driver has entered a subconscious state and stops the vehicle's DC motor accordingly.

\vspace{0.5cm}
\noindent\textit{Medication Management}\\
Adherence to medication schedules is vital but challenging for elderly individuals with memory issues. Fortunately, the integration of IoT offers a solution to this problem, and numerous research efforts have explored using IoT to track patients' medication compliance. In \cite{bharadwaj2017enhancing}, a medical box was created to remind individuals to take medications, featuring three trays for different times of the day. The system also measures vital health parameters (i.e., blood pressure, temperature, blood oxygen levels, etc.) and facilitates two-way communication between patients and doctors through a mobile application.



\vspace{0.5cm}
\noindent\textit{Wheelchair Management}\\
Wheelchairs are vital tools in the lives of individuals who are physically unable to move independently, providing both physiological assistance and psychological support. However, for individuals with brain damage who lack the capability to operate a wheelchair, researchers have been exploring the addition of navigation and tracking systems, with IoT playing a crucial role in these wheelchair advancements. An advanced and automated smart wheelchair was reported in a study \cite{kumar2020design}, which not only monitors movement but also offers features such as an umbrella, foot mat, head mat, and obstacle detection. These innovations have significantly improved interaction with the living environment and enhanced the user's overall experience.


\vspace{0.5cm}
\noindent\textit{Rehabilitation System}\\
The application of IoT in this field is versatile and has proven effective in various areas, including cancer treatment, sports injury recovery, stroke rehabilitation, and addressing physical disabilities. For instance, an innovative smart walker was introduced in a particular study \cite{nave2018smart}. Doctors and caregivers can access the collected data through a mobile application, facilitating better monitoring and support, as this walker utilizes a multi-modal sensor to monitor the patient's walking pattern.


\vspace{0.5cm}
\noindent\textit{Fitness}\\
Regular physical activity and maintaining a high level of fitness significantly influence the quality of an individual's life. Developers have created various applications leveraging the IoT to facilitate fitness monitoring and promote healthier lifestyles. These approaches include assessing users' activity levels and revealing metrics, including the duration of physical activity and periods of inactivity, by utilizing smartphone accelerometer data. Nowadays, wearable fitness trackers are readily available in the market and have gained popularity as convenient devices for monitoring fitness levels, for instance, smart mats to provide insights into users' workout routines \cite{sundholm2014smart} and fitness assessment and training load monitoring to optimize athletes' hydration strategies \cite{seshadri2019wearable}.



\vspace{0.5cm}
\noindent\textit{Other Notable Applications}\\
The application of IoT in the healthcare industry is incredibly diverse, extending far beyond the previously mentioned areas. There are numerous domains where IoT has already been implemented and where its potential benefits are being realized, leading to a significant increase in the adoption of healthcare IoT (HIoT) technology. For cancer treatment, IoT-based methods, for example, have emerged as powerful tools. An innovative IoT-based cancer treatment approach is introduced in a recent study that encompasses various stages, including chemotherapy and radiotherapy \cite{heshmat2018framework}. Additionally, this system securely stores lab test results on a cloud server, allowing physicians to monitor medication dosages and enabling remote consultations through a dedicated mobile application. Furthermore, HIoT has found applications in detecting skin lesions \cite{rodrigues2020new}, with notable advancements in lung cancer detection through state-of-the-art ML algorithms and IoT-based systems \cite{rajan2020fog, pradhan2020medical, liu2019deep}.

IoT has revolutionized the realm of surgical training and medical procedures by creating next-generation solutions. One such development involves a surgical training framework that employs virtual reality to simulate realistic training environments. This framework also enables interaction with surgeons from around the world, fostering collaborative learning and expertise sharing \cite{cecil2018iomt}. Monitoring haemoglobin levels in the blood has become more accessible through portable devices equipped with photoplethysmography sensors, light-emitting diodes (LEDs), and photodiodes. These devices enable the non-invasive measurement of haemoglobin levels, enhancing healthcare monitoring and diagnosis \cite{bhatia2019towards}.

Numerous other HIoT applications are currently in use or under research, underscoring the ongoing revolutionary impact of IoT in the field of healthcare. It is expected that the IoT will continue to drive advancements and improvements in healthcare delivery and patient outcomes.

\subsection{Smart Agriculture and Environment}
Agriculture holds a crucial position in a country's economic progress. Various factors, including soil moisture and environmental variables like carbon dioxide levels, temperature, and humidity, can significantly impact crop yields. To enhance agricultural outcomes, it becomes essential to implement robust surveillance systems in the fields. The integration of the IoT enables efficient achievement of this goal. The following section has explored several applications in this smart agricultural domain as well as a summary of the related research work in Table \ref{tab:agriculture}.

\begin{table}[ht]
\centering
\caption{Relevant studies in smart agriculture and environment domain}
\label{tab:agriculture}
\resizebox{\columnwidth}{!}{%
\begin{tabular}{l|l||l|l}
\hline
\rowcolor[gray]{.9}\textbf{Applications} & \textbf{Related Works} & \textbf{Applications} & \textbf{Related Works} \\ \hline \hline
Water-saving irrigation & \cite{yuanyuan2020research, li2017design} & Diseases Monitoring & \cite{chen2015study, lin2015self} \\ 
\rowcolor[gray]{.96}\begin{tabular}[c]{@{}l@{}}Animal and plant life\\  information monitoring\end{tabular} & \cite{xie2023deep, porto2011developing} & \begin{tabular}[c]{@{}l@{}}Intelligent agricultural\\  machinery\end{tabular} & \cite{sowjanya2017multipurpose, onishi2019automated} \\ 
\begin{tabular}[c]{@{}l@{}}Agricultural product quality\\  safety and traceability\end{tabular} & \cite{jiang2017research, gu2018construction} & \begin{tabular}[c]{@{}l@{}}Crop growth environment \\ monitoring\end{tabular} & \cite{materne2018iot, nagasubramanian2021ensemble} \\ \hline
\end{tabular}%
}
\end{table}


\vspace{0.5cm}
\noindent\textit{Water Saving Irrigation}\\
Water scarcity in agriculture is a growing concern, necessitating a dynamic irrigation approach due to varying crop water requirements. IoT integration revolutionizes traditional flood irrigation, offering a solution to water shortage problems in crop growth. Yang et al. \cite{yuanyuan2020research} have proposed an wireless sensor network based system leveraging neural networks for water-efficient irrigation. This method enhances irrigation efficiency by minimizing human intervention and reducing wastage due to excessive drainage.



\vspace{0.5cm}
\noindent\textit{Crop Growth Environment Monitoring}\\
Various environmental factors, including temperature, humidity, air pressure, carbon dioxide levels, soil temperature, and soil pH, play a crucial role in crop growth. IoT devices integrated into agricultural systems can sense and analyze these environmental factors, enabling remote field monitoring and the creation of an optimal farming environment tailored to these variables. Lin et al. \cite{lin2015self} have designed a wireless environmental monitoring system that harnesses soil energy to enable cost-effective remote monitoring of farmland environments.


\vspace{0.5cm}
\noindent\textit{Animal and Plant Life Information Monitoring}\\
Effective agricultural production requires comprehensive monitoring of both plant and animal information, which is crucial for enhancing production, increasing profitability, and ensuring high-quality product development.
\begin{enumerate}
    \item Animal Life Information Monitoring: Monitoring diverse aspects of animal behavior, including their food consumption, body temperature, activity levels, and health status, enables the tracking of their physiological and nutritional well-being, ensuring their healthy development. In \cite{xie2023deep}, the authors have proposed an infrared-based body temperature measurement system for pigs, allowing the early detection of diseases. 

\item Plant Life Information Monitoring: 
Wireless sensor devices, when connected to plants, enable remote and continuous monitoring of both external factors (such as diseases, pests, and leaf color) and internal factors (including chlorophyll content and photosynthetic rate). This technology allows for early disease detection and promotes overall healthy plant growth. Porto et al. \cite{porto2011developing} have introduced a citrus traceability system that assesses environmental conditions for optimal growth, identifying and preventing plant diseases to ensure robust citrus crop health.

\end{enumerate}

\vspace{0.5cm}
\noindent\textit{Intelligent Agricultural Machinery}\\
Intelligent machinery autonomously manages a wide range of agricultural operations, such as cultivation, sowing, transplanting, fertilization, pesticide application, feeding, irrigation, picking, and harvesting, all executed with precision and efficiency. Moreover, it has the capability to gather a variety of data about the farm, including soil moisture and water quality, as well as ambient information like temperature and humidity, which can be effectively harnessed to implement precision agriculture and enhance breeding practices \cite{ma2020intelligent}. IoT technology plays a key role in minimizing manual labor in agriculture by enabling remote monitoring and standardizing machinery functions through sensors and wireless communication \cite{ma2020intelligent}. Sowjanya et al. have introduced a versatile autonomous robot vehicle in \cite{sowjanya2017multipurpose} where the vehicle is equipped with Bluetooth technology for remote control and is capable of independently executing a range of tasks including farming, seeding, and irrigation. 


\vspace{0.5cm}
\noindent\textit{Agricultural Product Quality Safety and Traceability}\\
IoT significantly improves agricultural product quality, safety, and traceability, particularly in warehousing, logistics, and distribution, by enabling automatic identification, tracking, and accurate tallying of products. Various countries have implemented real-time traceability systems, such as the American, European, Swedish, Japanese, and Australian systems, recognizing the crucial need for effective tracking in agriculture. Jiang et al. \cite{jiang2017research} developed a comprehensive agricultural product safety traceability platform, facilitating real-time automatic data collection, processing, and display to enhance traceability and reduce associated tracking costs.


 
\vspace{0.5cm}
\noindent\textit{Diseases Monitoring}\\
The integration of IoT for real-time and continuous monitoring offers farmers and relevant authorities the capability to identify diseases at an early stage and implement preventive measures before they escalate. The farm's environment plays a pivotal role in disease occurrence. For example, the framework introduced in \cite{materne2018iot} integrates various sensor devices through wireless sensor networks for monitoring various environmental factors.

\subsection{Smart Home (SH)}
In a smart home, various types of sensors are strategically deployed, each with its own specific function. Smart homes simplify daily tasks for users, proving especially beneficial for those prone to forgetting routine actions like locking doors or turning off appliances. From smart door locks to the maintenance of household items like coffee machines, heaters, and smart bulbs, and even the use of surveillance cameras for enhanced security, smart homes offer a wide range of possibilities. Furthermore, users can control these devices through voice commands and remotely monitor their home equipment. Smart homes contribute to improved energy efficiency by automatically turning off devices not in use and notifying users of any unusual incidents. MavHome \cite{cook2003mavhome}, for example, employs prediction algorithms to perform various tasks in response to user-initiated events. As for energy conservation, an intelligent home achieves it through the utilization of sensors and the context-aware capabilities of IoT. Data gathered by these sensors is transmitted to a context aggregator, which then forwards the data to a context-aware service engine. This engine analyzes the data and determines appropriate actions. For instance, it may decide to turn off the air conditioning if the temperature is too cold, shut off the gas supply in case of a detected leak, or switch off the lights when there are no occupants at home \cite{han2010design}.


\subsection{Smart Manufacturing System (SMS)}
With the development and evolution of IoT, industrial IoT, artificial intelligence, and cyber-physical systems, many countries have opted to transform their manufacturing systems into smart manufacturing systems. Through the integration of smart technologies, these systems facilitate a rapid and extensive flow of data within and among manufacturing processes. Equipped with this data and employing advanced information and communication technology, smart manufacturing systems possess the capability to swiftly respond to global demands, efficiently utilise materials, energy, and labour resources, and deliver customised products on time \cite{qu2019smart}. What sets the smart manufacturing model apart from other manufacturing paradigms is its vision of the next generation of manufacturing with enhanced capabilities \cite{lu2016current}. These systems adapt to new circumstances by leveraging real-time information for intelligent decision-making and by proactively predicting and preventing potential failures.

\subsection{Internet of Robotics Things (IoRT)}
The Internet of Robotic Things is a concept that combines the principles of IoT and robotics. IoRT represents an emerging technology that incorporates robots within an IoT ecosystem as objects, enabling communication, collaboration, and automation. These robots seamlessly integrate into smart environments, performing a wide range of tasks. These tasks span from personal activities within smart homes to applications in the healthcare industry. Furthermore, they extend to professional activities such as monitoring, delivery, and object control within manufacturing industries or warehouses.

\subsection{Oil and Gas Industry}



The oil and gas industry, facing substantial costs and safety risks, embraces IoT-based remote monitoring for real-time field equipment oversight and data-driven decision-making. These solutions allow for the remote monitoring of field equipment, the analysis of field data, collaborative data-driven decision-making, and the implementation of control commands to optimize asset performance while mitigating health, safety, and environmental (HSE) hazards \cite{alsaadoun2019cybersecurity}. Furthermore, IoT integration in the oil industry focuses on reducing human labor, minimizing time wastage, and improving accuracy through automation, as exemplified by Equinor's well optimization system in the Bakken oil field \cite{freeman2018enabling}. By deploying IoT devices and ML algorithms in around 50 wells, Equinor achieved a 33\% increase in oil production through optimized well operation and maintenance.

\subsection{Smart Retail}
The adoption of IoT in the retail industry has created a flexible environment that benefits both customers and sellers. This shift allows the entire retail sector to migrate from offline to online, enabling customers to independently conduct their shopping through self-service while facilitating smooth interactions between retailers and their customers. Furthermore, retailers can employ IoT technologies like RFID to monitor products and deploy sensors to collect customer data, which they can then utilize for analyzing customer buying behavior and enhancing business profitability \cite{liu2018smart, jayaram2017smart}. Additionally, customers have the option to make payments through online transactions and monitor their orders using online services \cite{liao2020mobile}.

\subsection{Industrial Internet of Things (IIoT)}
The IIoT holds significant potential, according to numerous market researchers, by serving as an extension of the IoT specifically customized for the industrial sector and its applications. It empowers industries and enterprises to enhance and optimize their operations by leveraging M2M communication, big data analytics, and ML. The scope of IIoT is extensive, encompassing a wide array of connected industrial devices and systems. Connected electric meters, wastewater systems, flow gauges, pipeline monitors, manufacturing robots, and various other types of industrial equipment and devices are included in this list \cite{munirathinam2020industry}. One notable application of IIoT is in the mining industry, where companies like CISCO have implemented IIoT solutions to improve safety and efficiency in underground mines. These solutions involve connecting people, tracking the locations of miners and vehicles, monitoring vehicle statuses, and automating building controls.

\subsection{Social Life and Entertainment}
Several applications have been developed to monitor and enhance human social activities, in addition to work or professional activities, as social life and entertainment are integral parts of a person's life. Portable devices such as mobile phones and tablets possess sensing capabilities and communication technologies that facilitate interactions between individuals. Integrating IoT into an individual's social life can contribute to emotion detection, community building, and emotional support. CircleSense \cite{liang2013circlesense} is an application that analyzes a person's social activities using various sensors to identify their social circle. It also tracks the person's location via location sensors and employs Bluetooth technology to identify people in proximity. Camy, an artificial pet dog, expresses affection and empathy through the use of effective computing technology. This technology analyzes multiple aspects of a person's behavior, such as facial expressions, speech, body gestures, hand movements, and sleep patterns, to identify and appropriately respond to their emotions \cite{row2014camy}.
A Table \ref{tab:otherTabs} has been provided for presenting information about relevant works in the aforementioned domains.

\begin{table}[ht]
\centering
\caption{Relevant works in other applications}
\label{tab:otherTabs}
\resizebox{\columnwidth}{!}{%
\begin{tabular}{l|l||l|l}
\hline
\rowcolor[gray]{.9}\textbf{Applications} & \textbf{Related Works} & \textbf{Applications} & \textbf{Related Works} \\ \hline \hline
\begin{tabular}[c]{@{}l@{}}Social Life and \\ Entertainment\end{tabular} & \cite{liang2013circlesense, row2014camy} & \begin{tabular}[c]{@{}l@{}}Smart manufacturing \\ system\end{tabular} & \cite{qu2019smart, lu2016current} \\ 
\rowcolor[gray]{.96}Internet of robotics things & \cite{saravanan2018voice, chen2018wearable} & Oil and gas & \cite{alsaadoun2019cybersecurity} \\ 
Smart Retail & \cite{liu2018smart, jayaram2017smart} & Industrial IoT & \cite{zhou2017industrial, javaid2021upgrading} \\ 
\rowcolor[gray]{.96}Smart home & \cite{cook2003mavhome, yu2012posture} &  &  \\ \hline
\end{tabular}%
}
\end{table}


\section{Challenges and Future Directions}
\label{sec:challenges}
The remarkable position that IoT now occupies in today's world was once only a dream a few years ago. Today, IoT has captivated the entire globe and continues to extend its reach into various domains. Nonetheless, IoT also grapples with numerous issues and challenges that pose hurdles to its seamless implementation and expansion. This section offers an in-depth exploration of the key challenges confronting the IoT.
\subsection{Broad and Open Research Challenges}
Broad and open research challenges refer to complex and critical challenges in the IoT system that do not have any straightforward or predefined solutions. These challenges require extensive research and exploration in order to propose efficient solutions. They are ``broad'' in the sense that they encompass a wide range of related issues and considerations, and they are ``open'' because they may not have clear-cut or definitive answers, leaving room for ongoing research, experimentation, and discovery. This subsection provides a comprehensive listing of challenges in the IoT domain.

\vspace{0.5cm}
\noindent\textit{Building Intelligent Environments based on IoT Paradigm}\\
Creating a smart environment requires a vast number of devices, sensors, and complementary technologies to facilitate their interconnectivity. Managing this large volume of objects constitutes the initial challenge in the realm of IoT and intelligent environments. Moreover, the substantial task of collecting, storing, and conducting efficient analyses on enormous amounts of data remains a significant concern and creates collision issues within the IoT framework \cite{kassab2020z}. Handling massive amounts of devices and data in the IoT requires. One approach is to employ decentralized systems instead of centralized ones, reducing the volume of data sent to the cloud for processing. Techniques like data filtering, compression, and load balancing can further minimize the size of the data. Utilizing IoT technologies with robust device management and maintenance capabilities is also beneficial. Additionally, leveraging big data technologies like Hadoop and Spark can efficiently handle the substantial IoT data volumes. This holistic approach ensures readiness for the expanding IoT landscape.

\vspace{0.5cm}
\noindent\textit{Privacy and Security Challenges of IoT Applications}\\
The heterogeneity of IoT devices and the diversity of various IoT applications result in various security and privacy issues. People are primarily concerned about potential privacy invasions and security threats when using these technological devices. More information is provided below.
\begin{enumerate}
    \item Security: Different layers of the IoT are vulnerable to various kinds of attacks based on the technologies and protocols used in these layers. According to \cite{lin2017survey}, IoT layers including \textit{perception layer}, \textit{network layer}, and \textit{application layer} face various security attacks. As the major role of the \textit{perception layer} is to collect data, the security challenges in this layer focus on falsifying the data and destroying perception devices. Attacks including \textit{Node Capture Attacks, Malicious code Injection Attacks, False Data Injection Attacks, Replay Attacks, Cryptanalysis Attacks and Side Channel Attacks, Eavesdropping and Interference, and Sleep Deprivation Attacks} are faced by this layer \cite{lin2017survey}.

   Since the basic function of \textit{network layer} is to transmit collected data, particularly using wireless technologies, security challenges in this layer revolve around the availability of network resources and the wireless network. Challenges at the Network Layer encompass \textit{Denial-of-Service (DoS) Attacks, Spoofing Attacks, Sinkhole Attacks, Wormhole Attacks, Man-in-the-Middle Attacks, Routing Information Attacks, Sybil Attacks, and Unauthorized Access} \cite{lin2017survey}. \textit{Application layer}, which focuses on providing user-requested services, challenges primarily revolve around \textit{software attacks, including phishing attacks, malicious viruses/worms, malicious scripts} \cite{lin2017survey}.

  It is imperative to propose secure, robust, and reliable authentication schemes to detect and defend unauthorized access \cite{das2023lightweight}. Virus detection techniques and script detection techniques such as honeypot techniques, static code analysis, and dynamic action detection must be implemented to defend against worms, viruses, and malicious scripts bridging firewalls. Furthermore, introducing secured routing protocols is essential to ensure secure routing.

   \item Privacy: IoT devices continuously generate vast amounts of real-time data, which undergoes three main stages: (1) Data Collection, (2) Data Aggregation , and (3) Data Mining and Analytics. While these processes enhance our lives by providing various services, they also raise concerns about data privacy in the IoT. Privacy breaches in IoT can have serious repercussions for both the IoT network and its users, including financial losses, property damage, and even risks to human safety and security \cite{lin2017survey, singh2024dnacds}.

For instance, consider the smart grid, where adversaries can readily seize control of the smart metres, allowing them to access or manipulate the collected data. This could potentially compromise the confidentiality and privacy of energy consumption data. By using this altered data, utility providers may make inaccurate assessments of energy supply and demand within the grid, resulting in erroneous energy dispatch decisions. This, in turn, could lead to imbalances in energy supply and demand, potentially causing widespread power outages. In the healthcare industry, if an adversary manages to acquire a patient's health data, they could manipulate medication prescriptions or medical records, leading to significant health risks and potential insurance fraud. Therefore, it is imperative to deploy privacy preservation schemes to prevent data leakage and ensure that private data remains inaccessible to adversaries \cite{lin2017survey, namasudra2018security}.

There are three main groups for categorizing privacy-preserving mechanisms in the context of IoT data processing: (i) privacy preservation during data collection, (ii) privacy preservation during data aggregation, and (iii) privacy preservation during data mining and analytics. While various techniques, such as encryption and key management, can be applied to protect privacy in data collection, mining, and analytics, the majority of efforts in IoT privacy preservation have focused on data aggregation. Data aggregation involves processing relevant data in multiple locations, making it challenging to ensure privacy using traditional encryption methods. As a result, researchers have developed several privacy-preserving mechanisms specifically for data aggregation, which can be categorized as follows:
\begin{enumerate}
    \item  \textit{Anonymity-based privacy preservation}, which employs techniques like K-anonymity, L-diversity, and T-closeness to protect the privacy of identification information during data aggregation.

\item  \textit{Encryption-based privacy preservation}, which prevents adversaries from eavesdropping on data during aggregation by utilizing encryption techniques such as homomorphic encryption, commitment mechanisms, secret sharing, and zero-knowledge proofs \cite{das2022novel, das2023macpabe}.

\item  \textit{Perturbation-based privacy preservation}, where techniques such as data customization, data sharing, and random noise injection perturb raw data to ensure privacy during aggregation.
\end{enumerate}

Among these, perturbation-based privacy-preserving schemes are popular in IoT due to their direct operation on raw data. However, many of these perturbation-based privacy-preserving schemes sacrifice data utility to achieve privacy. This reduction in data utility can hinder the support of services requested by IoT applications. Therefore, a significant challenge in the field of data privacy preservation in the IoT is designing schemes that strike a balance between privacy and data utility, making it a crucial area for future research \cite{saifuzzaman2022systematic, lin2017survey}. In summary, safeguarding data privacy in the IoT is crucial to preventing these adverse outcomes and maintaining the security and integrity of both individuals and the IoT ecosystem.
\end{enumerate}

\vspace{0.5cm}
\noindent\textit{Compatibility}\\
Interconnecting devices from various vendors in an IoT network can pose monitoring and management challenges. Different industries currently rely on a multitude of standards to support their applications. Given the vast amounts of data, diverse device types, and the presence of various entities, utilising standard interfaces becomes crucial. This importance is amplified, especially for applications that need to accommodate both cross-organisational collaborations and a wide array of system limitations. Addressing these issues requires all industries to adhere to specific standards, but achieving such universal compliance can be a daunting and impractical task.

\vspace{0.5cm}
\noindent\textit{Scalability}\\
In the future, heterogeneous devices are expected to continuously join the ever-expanding IoT network. As a result, as the number of devices increases, ensuring smooth connectivity, effective data management, and overall system performance on a small scale becomes increasingly challenging. Therefore, the scalability of IoT poses an ongoing challenge for the future of this technology. To effectively address scalability challenges, it is essential to construct a scalable architecture, utilizing technologies like modular components, load balancers, and distributed systems.

\vspace{0.5cm}
\noindent\textit{Energy Efﬁciency}\\
Small smart devices that comprise IoT systems often have limited battery power, which is not easily replaceable. This limitation can lead to a global energy crisis and high power consumption, as well as constraints on memory and processing capabilities. Consequently, routing processes and compute-intensive applications may not run efficiently on these devices. While some routing protocols do support low-power communication, they are still in the early stages of development, and the constrained energy of smart devices may not be sufficient to fully utilize these WSN routing protocols. To tackle these challenges effectively, it's essential to emphasize the creation of low-power hardware and the adoption of energy-efficient protocols like MQTT-SN or CoAP rather than relying on more power-intensive alternatives such as HTTP. Additionally, harnessing over-the-air (OTA) firmware updates can ensure devices remain optimized and bug-free, thus diminishing the necessity for physical maintenance visits. Duty cycling, as another viable approach, aids in curbing power consumption significantly.

\vspace{0.5cm}
\noindent\textit{Mobility Management}\\
Mobility management in the IoT refers to the ability to handle devices that move within the network seamlessly. It is a crucial aspect because many IoT devices are not stationary and need to communicate as they change locations. The presence of mobile devices in IoT setups can lead to challenges in how routing protocols and IoT networks work efficiently. The current methods used for devices that move, like in sensor networks, mobile adhoc networks, and vehicular networks, can't effectively handle the various problems related to routing because these sensors have limited processing power and energy resources. To address these challenges, IoT systems employ various mobility management techniques and protocols, aiming to provide reliable and seamless communication for mobile devices in the IoT ecosystem.

\vspace{0.5cm}
\noindent\textit{Cost of Maintenance and Services}\\
The IoT network consists of a vast number of devices, utilizing various costly communication technologies. This inevitably leads to increased maintenance and service costs for these numerous devices and connections. Hence, a significant challenge lies in addressing this issue by designing devices and sensors that demand minimal maintenance.

\vspace{0.5cm}
\noindent\textit{Internet Disconnection Problem}\\
The disruption of internet connectivity, which is central to IoT operations, results in inferior performance from IoT applications and a decline in service quality. Furthermore, restrictions on the number of devices that can concurrently interact with the base station limit user access to these services. This issue is especially problematic in remote or unreliable network settings, where sustaining a consistent internet connection proves challenging. Consequently, addressing the problem of internet disconnections in IoT is imperative to maintain the reliability and efficiency of IoT systems.

\vspace{0.5cm}
\noindent\textit{Processing, Analysis and Management of Data}\\
The procedure for processing, analysing, and managing data is tremendously challenging because of the heterogeneous nature of IoT devices and the large scale of data generation. Currently, most systems utilize the centralized cloud-based systems for performing computationally intensive tasks and delivering data. However, an ongoing concern revolves around the limitations of traditional cloud architectures when it comes to efficiently handling the vast amounts of data generated and utilized by IoT-enabled devices. Additionally, these architectures struggle to support the associated computational demands while also meeting precise timing constraints. To address this challenge, most systems are currently relying on existing solutions like mobile cloud computing and fog computing, both of which utilize edge processing \cite{hussein2019internet}.



\vspace{0.5cm}
\noindent\textit{Other Challenges}\\
In addition to the previously mentioned challenges, IoT technology faces several other issues. The widespread adoption of IoT devices and technology, coupled with our increasingly reliant lifestyles, has led to users becoming highly dependent on IoT applications. This reliance is particularly critical in healthcare, where patients heavily depend on healthcare applications. Moreover, IoT devices can sometimes unexpectedly interfere with human activities, resulting in unanticipated and autonomous behaviors. The IoT network introduces ambiguity, making it challenging to distinguish between physical and virtual devices and even humans due to the ease of transformation between these categories. Quality and traffic control have become more complex due to the miniaturization and huge number of IoT devices. Managing unique identifications for each IoT device is also a growing concern. Furthermore, the IoT goes beyond geographical boundaries, with applications like healthcare offering services internationally. Nations face challenges due to the global reach of IoT, as data generated within their borders can be collected and transmitted to service providers located anywhere in the world, giving rise to concerns regarding data privacy and jurisdiction. Addressing these multifaceted challenges will require careful consideration and international collaboration to ensure the effective and secure implementation of IoT technology.

\subsection{Ethical Considerations}
The term ``ethical issue'' pertains to a situation or quandary characterized by a clash of moral principles, values, or ethical norms \cite{tzafestas2018ethics}. The realm of IoT confronts these ethical quandaries, necessitating individuals or organizations to make challenging choices amidst conflicting interests. These decisions frequently revolve around determining what is ethically correct or incorrect. Figure \ref{fig:ethicalIssues} illustrates prevalent ethical dilemmas within the IoT domain.

\begin{figure}[h]
    \centering
    \includegraphics[width=0.6\textwidth]{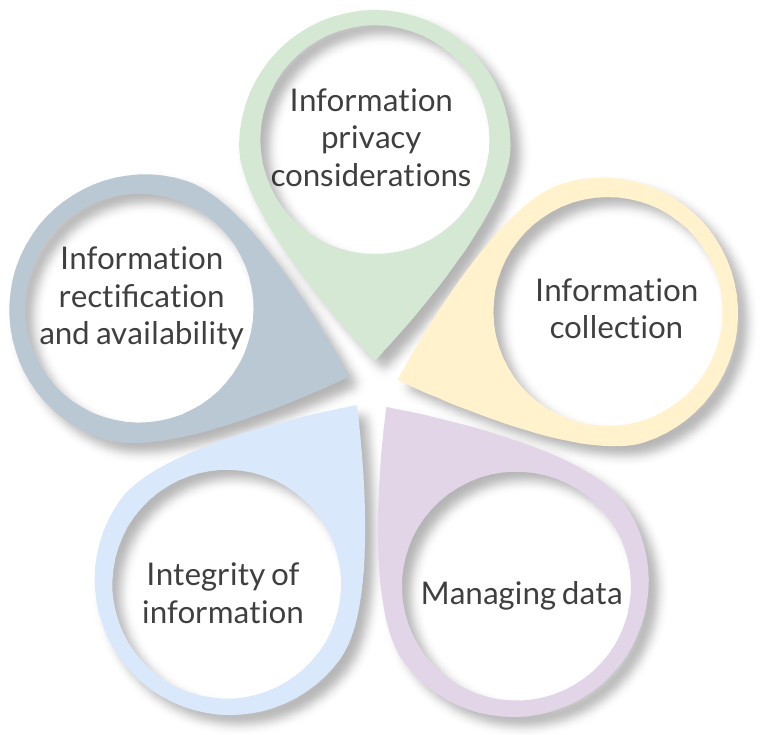}
    \caption{Ethical Issues in IoT}
    \label{fig:ethicalIssues}
\end{figure}

Adapted from \cite{carron2016internet}, the following outlines the five main categories that divide the ethical concerns related to IoT. The aim of putting forth these issues is to safeguard privacy rights by regulating how organizations manage information generated by IoT devices. These ethical standards, designed to establish guidelines for organizations, should also prompt individual concerns regarding privacy, as they serve as legal safeguards to protect individuals.
\begin{enumerate}
    \item \textit{Information privacy considerations}: 
   Organizations must handle the produced data with both openness and transparency. Except in specific cases, they should offer choices to individuals, allowing them not to disclose their identity or to use a pseudonym.
    \item \textit{Information collection}: 
    Organizations can gather requested data by implementing more stringent criteria for acquiring `sensitive' information. Conversely, they should specify their approach to handling unsolicited information. In both situations, organizations are required to delineate the circumstances surrounding the collection of this information and provide prior notification to relevant parties.
    \item \textit{Managing data}: 
    Organizations should specify the situations in which they might utilize or share the information they have gathered. Under specific conditions, an organization can employ personal data for direct marketing purposes. Nonetheless, they have the option of disclosing it internationally, but before doing so, they must set the safeguards that will be used to protect this information.

    \item \textit{Integrity of information}: 
    Entities should gather and share precise, current, and comprehensive information. Reasonable precautions must be in place to prevent misuse, interference, loss, and unauthorized access, changes, or disclosure.
    \item \textit{Information rectification and availability}: 
    When requesting access to their information, entities should clearly state their responsibilities for allowing access and making corrections to the information they possess. This involves the obligation to grant access and make necessary modifications, except in cases where a particular exception is applicable.
\end{enumerate}

\subsection{Legal and Regulatory Issues}
After identifying the challenges and ethical concerns, there is a need to consider the legal aspects concerning the effectiveness of current laws in safeguarding users within this context. The significance of this concern arises from the increasing blurring of the boundary between the physical and virtual realms in IoT. The ensuing questions, depicted in Fig. \ref{fig:legalIssues}, serve as examples of the issues that require discussion, as referenced in \cite{tzafestas2018ethics}.
\begin{figure}[]
    \centering
    \includegraphics[width=0.8\textwidth]{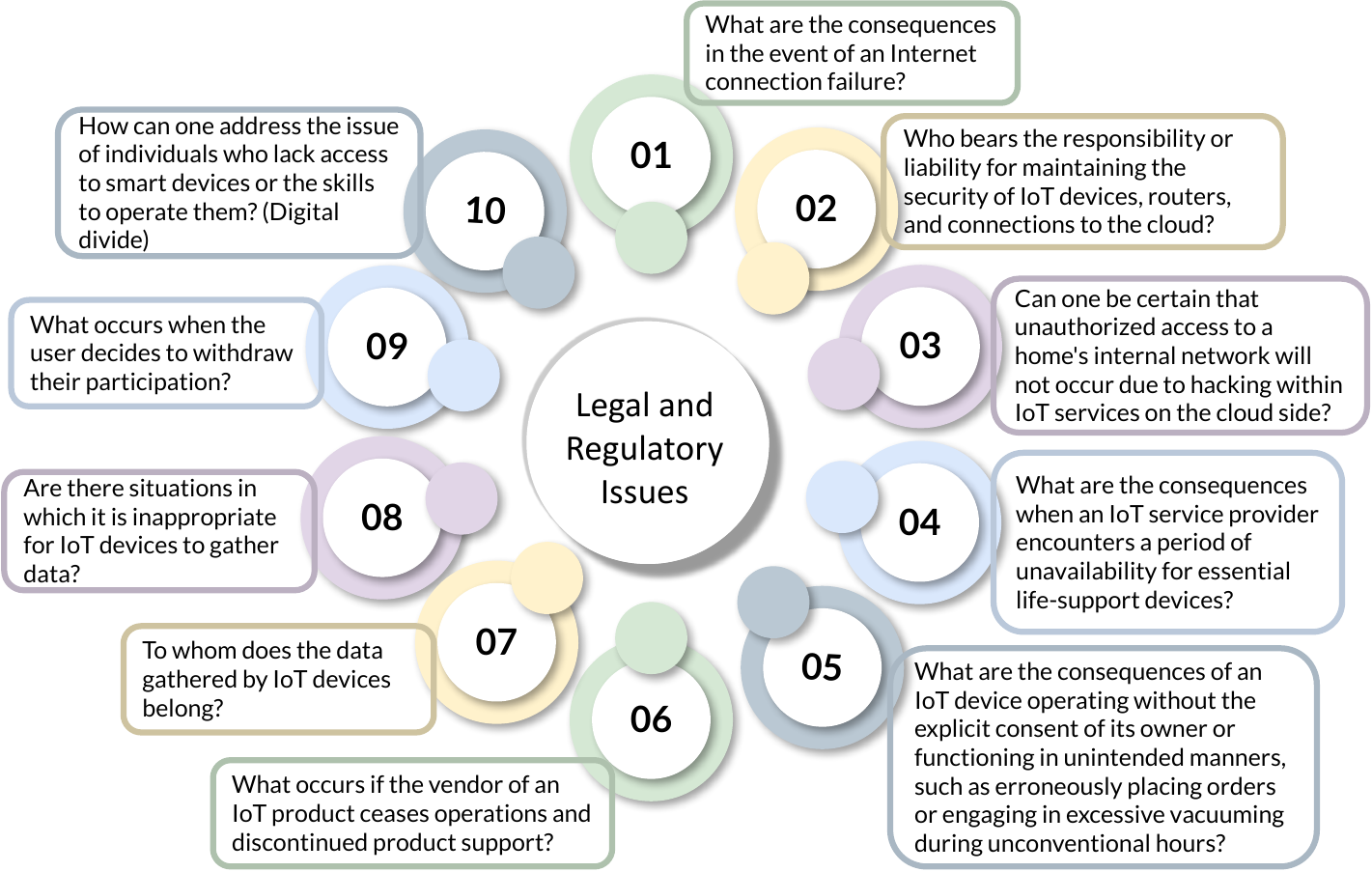}
    \caption{Legal and Regulatory Issues of IoT}
    \label{fig:legalIssues}
\end{figure}


Considering the challenges mentioned in Section \ref{sec:challenges}, it is evident that IoT faces several obstacles, including financial constraints, security vulnerabilities, and data privacy concerns, which can have life-threatening implications, especially in healthcare data breaches. To address these challenges effectively, a proactive approach involving extensive research is crucial. This research should focus on identifying IoT's specific issues, followed by the implementation of robust technical solutions. Effective execution of this process will not only ensure a secure IoT system but also encourage user trust in enrolling themselves within the IoT network.

To mitigate ethical concerns, raising user awareness is essential, along with the integration of self-adaptive security policies and dynamically modifiable policies during IoT application development. The introduction of new laws and standards is also necessary, integrating existing regulations like HIPPA, FIPPS, the Electronic Communication Privacy Act, and others to comprehensively address security, privacy, and legal issues. Additionally, addressing technical challenges involves the introduction of adaptable and new standards as well as the implementation of standard address identification. Some examples of these technical solutions include advanced encryption techniques, electronic signatures, the integration of standard protocols, and regulations to limit third-party data usage. This holistic approach aims to overcome the multifaceted challenges that the IoT faces.

\section{Conclusions}

The IoT has rapidly become an integral part of the 21\textsuperscript{st} century, enhancing daily decision-making and ushering in innovative consumer services like pay-as-you-use. The seamless integration of smart devices and automation technologies has revolutionized every aspect of our lives. However, amidst this technological marvel, we must acknowledge significant concerns related to security, privacy, intellectual property rights, safety, and trust. These concerns continue to demand further investigation. This chapter has provided a comprehensive overview of IoT for newcomers seeking to explore this domain and gain a thorough understanding to make future contributions.The chapter covers fundamental IoT concepts, historical development, architectures, advantages, and technology taxonomy. It explores diverse applications in domains such as smart cities and healthcare while addressing challenges and providing possible future directions. This discussion serves as a solid foundation for researchers interested in developing practical IoT projects or pioneering new theoretical approaches within the IoT field, equipping them with a deep understanding of various aspects of IoT. This, in turn, provides a good ground for researchers who are interested in designing realistic IoT projects or developing novel theoretical approaches in the IoT field by acquiring deep knowledge in different IoT aspects.

\begingroup
\let\clearpage\relax 

\bibliographystyle{ieeetr}
\bibliography{main}

\endgroup

\renewcommand\thesection{\Alph{section}} 
\clearpage



\end{document}